\documentclass[aps,prd,10pt,twocolumn,superscriptaddress,floatfix,nofootinbib,amsmath,amssymb,altaffilletter,preprintnumbers]{revtex4-2}
 
\usepackage{graphicx}
\usepackage{dcolumn}
\usepackage{bm}
\usepackage{braket}
\usepackage{dsfont, fontawesome}
\usepackage[colorlinks=true, pdfstartview=FitV, breaklinks=true]{hyperref}
\usepackage{color}
\usepackage[dvipsnames,table]{xcolor}
\hypersetup{urlcolor=BlueViolet, citecolor=Plum, linkcolor=PineGreen}
\usepackage{cleveref}
\newcommand{\grasp}{\texttt{GRASP}}
\newcommand{\ratip}{\texttt{RATIP}}
\newcommand{\bertha}{\texttt{BERTHA}}

\newcommand{\cdmpp}{\affiliation{ARC Centre of Excellence for Dark Matter Particle Physics,\\School of Physics, The University of Melbourne, Victoria 3010, Australia}}

\newcommand{\KCL}{\affiliation{Theoretical Particle Physics and Cosmology Group, Department of Physics, King's College London, UK}}
 
\newcommand{\unimelb}{\affiliation{School of Physics, The University of Melbourne, Victoria 3010, Australia}}

\begin{document}

\preprint{KCL-2022-30}

\title{Precise predictions and new insights for atomic ionisation from the Migdal effect}

\author{Peter Cox}
\email{peter.cox@unimelb.edu.au}
\cdmpp

\author{Matthew J. Dolan}
\email{dolan@unimelb.edu.au}
\cdmpp

\author{Christopher McCabe}
\email{christopher.mccabe@kcl.ac.uk}
\KCL

\author{Harry M. Quiney}
\email{quiney@unimelb.edu.au}
\unimelb


\begin{abstract}
The scattering of neutral particles by an atomic nucleus can lead to electronic ionisation and excitation through a process known as the Migdal effect. We revisit and improve upon previous calculations of the Migdal effect, using the Dirac-Hartree-Fock method to calculate the atomic wavefunctions. Our methods do not rely on the use of the dipole approximation, allowing us to present robust results for higher nuclear recoil velocities than was previously possible. Our calculations provide the theoretical foundations for future measurements of the Migdal effect using neutron sources, and searches for dark matter in direct detection experiments. We show that multiple ionisation must be taken into account in experiments with fast neutrons, and derive the semi-inclusive probability for processes that yield a hard electron above a defined energy threshold. We present results for the noble elements up to and including xenon, as well as carbon, fluorine, silicon and germanium. The transition probabilities from our calculations are publicly available \href{https://petercox.github.io/Migdal}{\faGithub}.
\end{abstract}


\maketitle

\section{Introduction}
\label{sec:intro}

Despite decades of searching, the precise nature of particle dark matter (DM) remains an enduring mystery. There is a wide-ranging program of direct detection experiments dedicated to measuring the properties of astrophysical DM in terrestrial laboratories~\cite{Billard:2021uyg}. These are based primarily on the possibility that DM scatters and imparts an $\mathcal{O}(\mathrm{keV})$ kinetic energy to an atomic nucleus. However, for DM masses less than a few GeV this method loses sensitivity, since the nuclear recoil energy becomes smaller than the experimental energy threshold.

An alternative approach for DM direct detection is to search for electromagnetic signals that may be produced when the DM interacts with the atomic nucleus~\cite{Vergados:2004bm, Moustakidis:2005gx, Ejiri:2005aj, Bernabei:2007jz}. The possibility that an electron may be emitted from an atom after the sudden perturbation of the nucleus has been known since the early 1940s~\cite{Migdal:1939, Migdal:1941, Feinberg:1941, Migdal:1977} and has become known as the `Migdal effect' within the DM community~\cite{Bernabei:2007jz}. The broader importance of the Migdal effect for direct detection searches has only recently been established in~\cite{Ibe:2017yqa, Dolan:2017xbu}, and further studied in~\cite{Sharma:2017fmo, Essig:2019xkx, Bell:2019egg, Baxter:2019pnz, Knapen:2020aky, GrillidiCortona:2020owp, Liu:2020pat, Liang:2020ryg, Bell:2021zkr, Knapen:2021bwg, Acevedo:2021kly,Blanco:2022pkt}. Although the production of electromagnetic signals is suppressed relative to the rate of conventional elastic nuclear scattering, there is a window for sub-GeV DM where the nuclear recoil energy falls below threshold while the electromagnetic signal remains observable.\footnote{Polarisation (or atomic) bremsstrahlung produces a similar effect~\cite{Amusia1987, Amusia1988, Korol2014, Kouvaris:2016afs, McCabe:2017rln}, but its rate is suppressed relative to the Migdal effect~\cite{Bell:2019egg}.} Several experiments have now exploited this to constrain the sub-GeV DM parameter space~\cite{LUX:2018akb, EDELWEISS:2019vjv, CDEX:2019hzn, XENON:2019zpr, SENSEI:2020dpa, COSINE-100:2021poy, EDELWEISS:2022ktt, DarkSide-50:2022ugn}.

Despite the importance of the Migdal effect for DM searches, the emission of an electron after a sudden jolt to the nucleus by an electrically-neutral projectile has not been measured experimentally.\footnote{The Migdal effect has been measured experimentally in related scenarios where the recoil is due to $\alpha$-decay~\cite{Rapaport1975, Rapaport1975-2, Fischbeck1975, Fischbeck1977} or $\beta$-decay~\cite{Boehm1954, Berlovich1965, Couratin2012, Lienard2015, Fabian2018}.} This has motivated several experimental proposals that aim to systematically study the Migdal effect over a wide range of energies and with different atomic species~\cite{Nakamura:2020kex, Bell:2021ihi, MIGDALa}. The proposals follow the standard practice within the DM community of using neutrons as the electrically-neutral proxy for DM and cover a range of neutron energies from 17\,keV in Ref.~\cite{Bell:2021ihi}, 565\,keV in Ref.~\cite{Nakamura:2020kex}, to 2.5\,MeV and 14.7\,MeV in the MIGDAL experiment~\cite{MIGDALa}. 

The characteristic signal of the Migdal effect is a recoiling ion and an ionisation electron emerging from a common vertex. While the use of lower-energy neutrons allows the Migdal effect to be studied in the kinematic regime relevant for DM experiments, it has the disadvantage that the nuclear recoil and ionisation electron cannot be separately resolved, either spatially or energetically; precise modelling of nuclear recoil quenching and the detector response is then required to test the Migdal effect (see e.g.,~\cite{Collar:2021fcl, Liao:2021yog}). In contrast, higher-energy neutrons probe a different energy regime from DM experiments, yet offer the possibility of indirectly or directly imaging the Migdal effect: indirectly in high-pressure gas, where the Migdal effect followed by de-excitation of the atom induces a `two-cluster' topology~\cite{Nakamura:2020kex}; and directly in low-pressure gas, where nuclear recoil and ionisation electron tracks can be imaged emerging from a common vertex~\cite{MIGDALa}.

\begin{figure}[t!]
    \includegraphics[width=0.9\columnwidth]{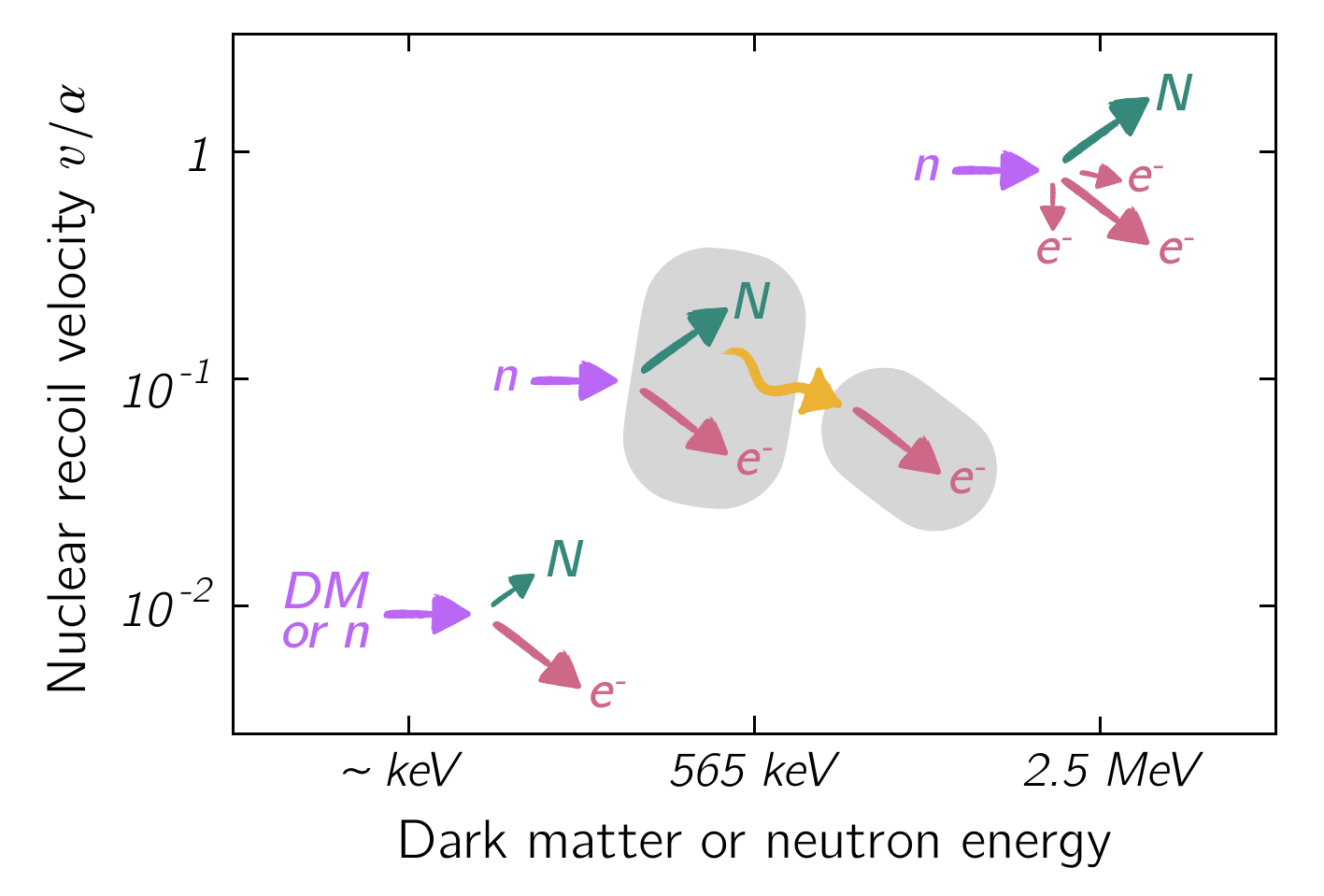}
    \caption{Schematic representation of the different regimes for DM- and neutron-induced Migdal processes. The dimensionless quantity $v/\alpha$ is the nuclear recoil velocity relative to the fine-structure constant ($c=1$). Bottom left: DM and low-energy neutron scattering, where the recoiling ion ($N$) is represented with a short arrow since its energy often falls below the detection threshold. Centre: Mid-energy neutron scattering in high-pressure gas can induce a distinctive two-cluster topology. The recoiling ion and the electron ionised through the Migdal effect induce one cluster, while the X-ray emitted following de-excitation of the recoiling ion induces a second cluster, separated by several~cm. Top right: High-energy neutron scattering from D-D/D-T fusion generators typically produces one hard and multiple soft ionisation electrons, as represented by one longer and multiple shorter electron arrows.}
    \label{fig:regimes}
\end{figure}

The qualitatively different scattering regimes for DM and neutron scattering are illustrated in \cref{fig:regimes}. This figure also introduces the dimensionless ratio $v/\alpha$, the magnitude of the nuclear recoil velocity relative to the fine-structure constant multiplied by the speed-of-light (we work in natural units with $\hbar=c=1$), which will serve as a key parameter. While DM direct detection experiments operate in the regime $v\ll \alpha$, neutron experiments can potentially test a much wider parameter space, up to $v\simeq \alpha$.

The purposes of this article are three-fold. Firstly, given the increasing interest in the Migdal effect for DM searches, and given that there is the potential for upcoming experiments to use this effect for DM discovery~\cite{Aalbers:2022dzr}, precise predictions are needed for the low-energy regime probed by DM experiments. The previous state-of-the-art calculations appeared in Ref.~\cite{Ibe:2017yqa} and used a relativistic self-consistent mean-field approach with the approximation of a local central potential~\cite{gu2008flexible}. We provide two additional, independent calculations of the Migdal effect. The first uses the \grasp~\cite{Jonsson:2007, Jonsson:2013, Fischer:2019} and \ratip~\cite{Fritzsche:2012} codes to calculate the bound and continuum wavefunctions, respectively. The second uses the \bertha~\cite{Quiney:1998} code, and serves primarily as a cross-check of our results. Both approaches employ the canonical Dirac-Hartree-Fock method, and include the full non-local exchange potential in place of the empirical local potential employed in~\cite{Ibe:2017yqa, gu2008flexible}. 

Secondly, the calculation of the Migdal effect in Ref.~\cite{Ibe:2017yqa} assumed the dipole approximation. While this is a good approximation in the regime $v\ll \alpha$ characteristic of DM experiments, it is expected to fail at the higher values of $v/\alpha$ that can be probed in neutron scattering. We therefore calculate the Migdal transition probabilities without making this approximation, instead evaluating the multi-electron matrix elements. Furthermore, we provide a detailed characterisation of the regime of validity of the dipole approximation.

Thirdly, multiple electrons can be ionised through the Migdal effect in the regime where $v\simeq \alpha$~\cite{Vegh_1983, wauters1997recoil, berakdar2001electronic, Talman:2006, liertzer2012multielectron, pindzola2014neutron, pindzola2020neutron}, and so we calculate the probability of double ionisation events in neutron-beam experiments, as well as in DM scattering. We also introduce the `semi-inclusive' probability to produce a hard electron in addition to one-or-more soft electrons (which would be below typical experimental thresholds and therefore not observable). We find that this semi-inclusive rate is needed to obtain accurate predictions for high-energy neutron beam experiments; this is due to the importance of multiple ionisation, which causes the semi-inclusive rate to grow with the energy of the recoiling nucleus. 

We present results for the noble elements from helium to xenon, as well as for carbon, silicon, germanium and fluorine. The noble elements are widely used in DM scattering experiments, with CYGNUS planning to operate with gaseous~He~\cite{Vahsen:2020pzb, Vahsen:2021gnb}, NEWS-G with gaseous~He and~Ne~\cite{NEWS-G:2022kon, Hamaide:2021hlp, NEWS-G:2023qwh}, while Darkside~\cite{DarkSide-20k:2017zyg} and DEAP-3600~\cite{DEAP-3600:2017ker} operate with liquid Ar, and LZ~\cite{LZ:2019sgr}, PANDA-X~\cite{PandaX:2014mem}, XENONnT~\cite{XENON:2020kmp}, and the proposed XLZD~\cite{DARWIN:2016hyl, Aalbers:2022dzr} experiment use liquid Xe. The MIGDAL experiment~\cite{MIGDALa} plans to operate with~C and~F in the form of low-pressure $\rm{CF}_4$ gas, as well as $\rm{CF}_4$ mixed with other gases including the noble elements, Si, and~Ge.\footnote{A number of DM experiments use~Si and~Ge {\it semiconductor} detectors: the relevant formalism for the Migdal effect in semiconductors has been derived in~\cite{Knapen:2020aky, Knapen:2021bwg, Liang:2022xbu}.} The Migdal transition probabilities we have calculated are publicly available~\cite{Git_results} for use by the community.

The paper is organised as follows. In Section~\ref{sec:derivation} we discuss the evaluation of the Migdal matrix element and introduce the semi-inclusive transition probability. In Section~\ref{sec:probabilities} we present our numerical results in the context of select illustrative examples. Section~\ref{sec:pheno} applies our results to DM and neutron-scattering experiments, and we conclude in Section~\ref{sec:conc}. Several appendices contain details pertinent to our calculations.

\section{Migdal Probability: generalities}
\label{sec:derivation}

There are several derivations of the Migdal effect in the literature~\cite{Migdal:1941, Lovesey:1982, Ibe:2017yqa}, all of which converge on the same result (discussion on this point can be found in~\cite{Kahn:2021ttr}). The most straightforward and intuitive way to derive the transition matrix element for an $N$-electron atom is through an argument due to A.~Migdal~\cite{Migdal:1941, Migdal:1977}. In the rest frame of the nucleus it is the electron cloud that is boosted due to the nuclear recoil. Final-state electronic wavefunctions in this frame are therefore obtained by applying a Galilean boost to the electronic wavefunctions of the atom at rest. The required matrix element is then given by 
\begin{equation} \label{eq:migdal_formula}
    \bigg\langle \Psi_f \bigg| \exp\bigg(im_e\bm{v}\cdot\sum_{k=1}^N \bm{r}_k\bigg)  \bigg|\Psi_i\bigg\rangle \,,
\end{equation}
where $m_e$ is the electron mass, $\bm{v}$ is the nuclear recoil velocity, and the sum is over the position operators $\bm{r}_k$ of the $N$ electrons. The initial and final state electronic wave functions of the atom in the $v=0$ frame are denoted by $\Psi_i$ and $\Psi_f$ respectively.

Although the above argument relies on the sudden/impulse approximation (the assumption that the projectile-nucleus interaction time-scale is short with respect to the electronic response time), the matrix element in \cref{eq:migdal_formula} holds in general, up to corrections of $\mathcal{O}\left( m_e/m_\mathrm{N} \right)$. If the interaction with the nucleus is long range, as in the case of dark matter scattering via a light mediator, then there is an additional form factor $F(q)\sim1/q^2$. The situation is more complicated if the projectile interacts with electrons, in which case there will be an atomic form factor~\cite{Vegh_1983}. While neutrons interact with electrons via a magnetic dipole interaction, this effect is estimated to be negligible~\cite{Migdal:1977,Talman:2006}.

\subsection{Exclusive transition probability}
\label{sec:exclusive}

We first consider the probability to transition to a specific final state, which we term the {\it exclusive} transition probability. The matrix element in \cref{eq:migdal_formula} contains an $N$-electron operator. As was pointed out in Ref.~\cite{Talman:2006} in the context of closed-shell atoms, this can be re-written in terms of single-electron matrix elements when the initial and final state wavefunctions are expressed as anti-symmetric products of single-electron wavefunctions. We denote the initial and final states~by 
\begin{align}
    \big|\Psi_i\,\big\rangle &= \big|\psi_{a_1}\psi_{a_2}\ldots\psi_{a_N}\big\rangle \,, \notag \\
    \big|\Psi_f\big\rangle &= \big|\chi_{b_1}\chi_{b_2}\ldots\chi_{b_N}\big\rangle \,,
\end{align}
where \{$\ket{\psi_a}$\} and \{$\ket{\chi_b}$\} are two orthonormal bases of four-spinor single-electron wavefunctions from which the initial and final wavefunctions are constructed, respectively. The subscripts $a_i$ and $b_i$ denote the set of quantum numbers that describe the wavefunction; for relativistic bound states these are $n_i$ ($E_i$ for continuum states), $\kappa_i$, and $m_i$ (see~\cref{app:relativistic-MEs}). The transition matrix element for the Migdal effect then simplifies to
\begin{equation} \label{eq:migdal_det}
    \big\langle \Psi_f \big| e^{im_e\bm{v}\cdot\sum_k \bm{r}_k} \big|\Psi_i\big\rangle = \det(M) \,,
\end{equation}
with the $N\times N$ matrix of single-electron matrix elements
\begin{equation} \label{eq:migdal_matrix}
    M_{\beta\alpha} = \big\langle\chi_{b_\beta}\big| e^{im_e\bm{v}\cdot\bm{r}} \big|\psi_{a_\alpha}\big\rangle \,.
\end{equation}
Following the standard approach, we evaluate these matrix elements by expanding the exponential operator in spherical tensors, as discussed in detail in~\cref{app:relativistic-MEs}. 

The exclusive transition probability is then
\begin{equation} \label{eq:probDet}
    p_v\left(\big|\Psi_i\big\rangle \to \big|\Psi_f\big\rangle \right) = \det(M M^{\dagger}).
\end{equation}
In practice, the relevant initial state is the atomic ground state, with $\ket{\psi_{a_\alpha}}$ the occupied orbitals in the ground state. On the other hand, for excitation (ionisation) processes $ \ket{\Psi_f}$ will include one or more excited (continuum) orbitals. 

In this section we have, for clarity of presentation, considered the case where the atomic wavefunction can be expressed as a single Slater determinant. However, eigenstates of the atomic Hamiltonian are more accurately represented by configuration state functions (CSFs), which are linear combinations of Slater determinants. The generalisation of \cref{eq:migdal_det} and~\cref{eq:probDet} to this case is provided in \cref{app:open-shell}.

\subsection{Semi-inclusive transition probability}
\label{sec:inclusive}

The probability for multiple ionisation becomes significant when the recoil velocity $v\gtrsim\alpha$, as we shall see in explicit examples in section~\ref{sec:probabilities}. However, for atoms with more than a handful of electrons, the large number of possible final states makes it impractical to calculate all of the exclusive transition probabilities. This motivates us to introduce the \emph{semi-inclusive} probability $p_v(\ket{\Psi_i} \to \ket{\chi_{b_1} X_\text{soft}})$, which includes all final states with an electron in the state $\ket{\chi_{b_1}}$ and the remaining electrons, denoted collectively by $X_\text{soft}$, in either bound or continuum states with energies below some threshold $E_\text{th}$.

The semi-inclusive rate is of particular relevance for experiments aiming to observe the Migdal effect in neutron scattering. This is because the differential probability falls rapidly with increasing ionisation electron energy; hence, in an experiment where the electron energy threshold is $E_\text{th}\sim\mathcal{O}(\mathrm{keV})$, the expected signal is one hard electron with additional sub-threshold excited or ionisation electrons.

The semi-inclusive transition probability is
\begin{multline} \label{eq:inclusive_definition}
    p_v(\ket{\Psi_i} \to \ket{\chi_{b_1} X_\text{soft}}) = \\
    \frac{1}{(N-1)!}\sum_{b_2, \ldots, b_N}^{(E<E_\text{th})} \bigg| \Big\langle\chi_{b_1}\ldots\chi_{b_N}\Big| e^{im_e\bm{v}\cdot\sum_k \bm{r}_k} \Big|\Psi_i\Big\rangle \bigg|^2 \,,
\end{multline}
where the sum is over all states where $N-1$ electrons have energy less than $E_\text{th}$ (here and in the following, the sum should be understood to include both the sum over bound orbitals and the integral over continuum orbitals). For $N>2$ it is impractical to directly evaluate this expression but, given that the probability of producing multiple electrons above an $\mathcal{O}(\rm{keV})$ threshold is negligible (justified below), a good approximation to the semi-inclusive probability is obtained by replacing $\sum_{b_2,\ldots,b_N}^{(E<E_{\mathrm{th}})} \to \sum_{b_2,\ldots,b_N}^\text{all states}$. In other words, the semi-inclusive probability is approximately equal to the probability of producing one hard electron with additional hard or soft electrons, $p_v(\ket{\Psi_i} \to \ket{\chi_{b_1} X_\text{soft}}) \approx p_v(\ket{\Psi_i} \to \ket{\chi_{b_1} X_\text{all}})$. This leads to \newline
\begin{widetext}
\begin{align} \label{eq:inclusive}
    p_v(\ket{\Psi_i} \to \ket{\chi_{b_1} X_\text{soft}}) &\approx \Big\langle\Psi_i\Big| e^{-im_e\bm{v}\cdot\sum_k \bm{r}_k} \bigg( \frac{1}{(N-1)!} \sum_{b_2, \ldots, b_N}^\text{all states} \Big|\chi_{b_1}\ldots\chi_{b_N} \Big\rangle \Big\langle \chi_{b_1}\ldots\chi_{b_N}\Big| \bigg) e^{im_e\bm{v}\cdot\sum_k \bm{r}_k} \Big|\Psi_i\Big\rangle \notag \\
    &= \sum_{\alpha=1}^N \Big| \big\langle\chi_{b_1}\big| e^{im_e\bm{v}\cdot\bm{r}} \big|\psi_{a_\alpha}\big\rangle \Big|^2 \,,
\end{align}
\end{widetext}
where the final result contains only single-electron matrix elements, with the sum over the occupied orbitals in the initial state. In going from the first to the second line of \cref{eq:inclusive} we have used the orthogonality and completeness of the~\{$\ket{\chi_b}$\}.

In deriving \cref{eq:inclusive}, we assumed that the probability of producing multiple electrons above threshold is negligible. We verify this numerically for several atoms in \cref{sec:probabilities}, but it is expected to be true in general for sufficiently large $E_{\mathrm{th}}$. This is because the high-energy continuum radial wavefunctions oscillate rapidly, with wavenumber $\sim\sqrt{2m_e E_e}$, which suppresses the radial integral with the initial-state bound electrons in the single-electron matrix elements (see \cref{eq:1e-ME}). However, this suppression disappears at high recoil velocities where the spherical Bessel function $j_L(m_evr)$ in the integral oscillates with a comparable wavenumber. We therefore expect the approximation in \cref{eq:inclusive} to eventually break down when $v\gtrsim\sqrt{2E_\text{th}/m_e}$ and the probability for multiple hard emission becomes significant. For a typical experimental threshold, this corresponds to $v/\alpha \gtrsim 8.6\sqrt{(E_{\rm{th}}/1~\mathrm{keV})}$, which is larger than the maximum recoil velocity with a D-T neutron source (where the neutron energy is $\sim 14\,\mathrm{MeV}$) for every element except helium. We have confirmed numerically for helium that this expression provides an excellent estimate for $\sim$keV thresholds. 

Note that the above derivation can also be straightforwardly extended to obtain, for example, the two electron semi-inclusive rate $p(\ket{\Psi_i} \to \ket{\chi_{b_1} \chi_{b_2} X_\text{soft}})$. However, this is unlikely to be of practical relevance due to the very low probability of producing two high-energy ionisation electrons. The generalisation of \cref{eq:inclusive} to open-shell systems is provided in \cref{app:open-shell}.

\subsection{Dipole approximation}
\label{sec:dipole-approx}

Most previous works have evaluated the matrix element in~\cref{eq:migdal_formula} using the dipole approximation, where the exponential is expanded to first order in $v$, 
\begin{equation}
    \exp\left(im_e\bm{v}\cdot\sum_{k=1}^N \bm{r}_k\right) \approx 1 + im_e\bm{v}\cdot\sum_{k=1}^N \bm{r}_k + \dots \,.
    \label{eq:dipole}
\end{equation}
It is important to establish the regime of validity of this approximation. A simple estimate uses the fact that atomic wavefunctions have support on distances of order the Bohr radius, $a_0$; the relevant expansion parameter is then $m_e v a_0=v/\alpha$. It has previously been argued~\cite{Liu:2020pat} that for ionisation from inner shells $a_0$ should be replaced by the effective Bohr radius $a_0/Z_n$, where $Z_n$ is the effective nuclear charge for the given shell. However, for the exclusive single ionisation rate, this argument is incorrect. There is an additional subtlety, which is that the dominant correction to the dipole approximation comes not from the bound-continuum matrix element, but from outer shell bound-bound matrix elements appearing in the determinant of \cref{eq:migdal_matrix}.

The breakdown of the dipole approximation can be understood using hydrogenic wavefunctions. The leading correction comes from the $ns \to np$ matrix element, which at leading order in $v$ is proportional to $v n^2/(\alpha Z_n)$~\cite{Bethe1957}. The dipole approximation is therefore expected to be valid only when $v \ll \alpha Z_n/n^2$, where $n$ should be taken as the principal quantum number of the valence shell. This means that the dipole approximation generally fails at lower~$v$ for larger atoms. While the above argument relies on hydrogenic wavefunctions, the same behaviour is found in our numerical results, as we discuss in \cref{sec:dipole-validity}.

Interestingly, the dipole expansion provides a much better approximation to the semi-inclusive rate. This is because the latter depends only on the bound-continuum matrix elements in \cref{eq:inclusive} and, for sufficiently high $E_\text{th}$, inner shell ionisations provide the dominant contribution. The dipole approximation to \cref{eq:inclusive} therefore holds when $v\ll Z_n/(m_e a_0)=\alpha Z_n$, where $Z_n$ here denotes the effective charge for the shell that gives the largest contribution. It is interesting that despite the dipole operator only allowing for single electron transitions, it provides a good approximation to the \emph{semi-inclusive} probability, including multiple ionisation, up to corrections of order $(v/(\alpha Z_n))^2$.

\section{Migdal Probability: Numerical results and illustrative examples}
\label{sec:probabilities}

The previous section provided a general discussion of the Migdal effect and the methods for calculating transition probabilities when atomic wavefunctions are expressed as anti-symmetric products of single-electron wavefunctions. In this section, we describe our implementations of the general theory, and discuss select examples to illustrate particularly interesting aspects of our results. 

We calculate the atomic wavefunctions with two independent implementations of the canonical Dirac-Hartree-Fock (DHF) formalism (for a recent review, see~\cite{Fischer_2016}), which use either basis-set or finite-difference methods for the radial functions. The basis-set approach, as implemented in the \bertha~\cite{Quiney:1998} package, is ideally suited for the calculation of integrated probabilities, since the sum over states approaches completeness in a systematic way as the size of the basis set is increased. For the calculation of differential ionisation probabilities we use the \grasp~\cite{Jonsson:2007, Jonsson:2013, Fischer:2019} and \ratip~\cite{Fritzsche:2012} packages to calculate the bound and continuum wavefunctions, respectively. Further details of our atomic calculations are provided in \cref{app:grasp,app:bertha}. The use of two completely independent implementations also allows for important cross-checks of our results, and we find excellent agreement between the two approaches. The outputs from our numerical calculations are tables of transition probabilities, which we utilise in the subsequent discussion, and are made available for use by the community~\cite{Git_results}.

\begin{figure}[t!]
    \includegraphics[width=0.95\columnwidth]{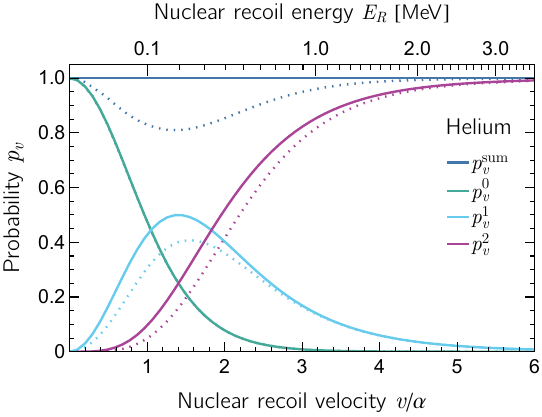}
    \par\vspace{4ex}{\par}
    \includegraphics[width=0.95\columnwidth]{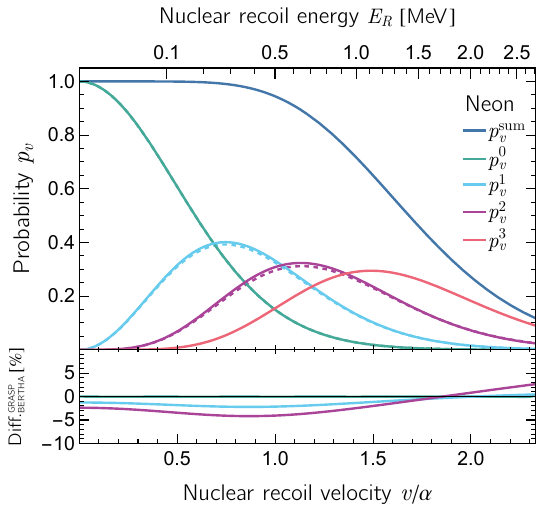}
    \caption{Integrated transition probabilities as a function of nuclear recoil velocity for helium (upper) and neon (lower). The solid lines show the probability for no electronic transition $p_v^0$, single transition $p_v^1$, double transition $p_v^2$, and, for neon, triple transition $p_v^3$. Upper panel: The dotted lines correspond to single (cyan) and double (purple) ionisation (i.e.\ transitions without bound excitations). Lower panel: The solid lines were calculated using \bertha\ and the dashed ones with \grasp/\ratip. The lower sub-panel shows the difference between them, $100\times(p_v^\grasp/p_v^\bertha-1)$.}
    \label{fig:integrated}
\end{figure}

\subsection{Integrated transition probabilities}

We begin by discussing the integrated transition probabilities for helium and neon, as plotted in \cref{fig:integrated} as a function of the nuclear recoil velocity. These atoms cleanly illustrate many of the general features seen in larger atoms. Helium and neon are also of interest as proposed target gases for both the NEWS-G direct DM and MIGDAL neutron-scattering experiments. 

In the upper panel of \cref{fig:integrated}, the solid lines show the ground state-to-ground state (green, $p_v^0$), single transition (cyan, $p_v^1$) and double transition (purple, $p_v^2$) probabilities, and their sum (dark blue, $p_v^\text{sum}$). These include all possible transitions to bound-excited-orbitals or ionised continuum states, integrated over electron energies. For low nuclear recoil velocities, the most probable outcome is that the entire atom recoils, remaining in the electronic ground state; on the other hand, for sufficiently high-velocity recoils the atom is always fully ionised. In the special case of helium, all of the above quantities can be expressed purely in terms of ground-state matrix elements, using the completeness of the single electron wavefunctions and conservation of probability. Our implementations with \bertha\ and \grasp/\ratip\ give near identical results for these matrix elements.

The dotted lines in the \cref{fig:integrated} upper panel show the probabilities for single ionisation (cyan) and double ionisation (purple) calculated with \grasp/\ratip; the dark blue-dotted line shows the sum of the ionisation probabilities and $p_v^0$. From the difference between the ionisation-only and single/double transition (solid) curves we see that, in helium, the final states with excited bound electrons contribute at most 20\% of the total probability, and are completely negligible at high recoil velocities. 

\begin{figure*}[t!]
    \includegraphics[width=0.95\columnwidth]{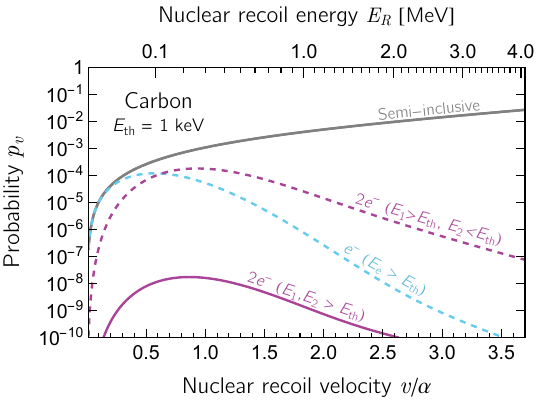}
    \hspace{1.0ex}
    \includegraphics[width=0.95\columnwidth]{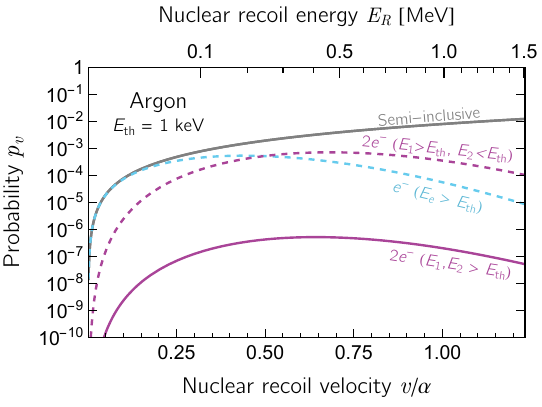}
    \caption{Semi-inclusive ionisation probabilities with $E_\text{th}=1\,$keV as a function of nuclear recoil velocity (solid grey). The dashed cyan and purple curves show the contributions from exclusive single and double transitions, respectively. The solid purple line shows the probability for double ionisation with both electron energies above 1\,keV. The panels are for carbon (left) and argon (right).}
    \label{fig:inclusive}
\end{figure*}

In the lower panel of \cref{fig:integrated} we show similar results for neon. Here, the solid lines were obtained using \bertha\ and we include up to triple transitions. The neon calculation illustrates an important general feature for many-electron atoms, which is the significant probability for multiple ionisation with increasing recoil velocity. We have not calculated quadruple and higher transitions, since it is computationally intensive, but it is clear that these become important in neon when $v\gtrsim\alpha$ and $p_v^\text{sum}=p_v^0+p_v^1+p_v^2+p_v^3$ (dark blue curve) falls significantly below one. Ultimately, for nuclear recoil velocities exceeding the orbital velocity of the inner shell electrons we expect the nucleus to effectively leave the entire electron cloud behind, leading to the complete ionisation of the atom. For neon this corresponds to a velocity of $v\gtrsim10\alpha$ or a recoil energy of around 50\,MeV; recalling that the binding energy per nucleon is around 8\,MeV for neon, it seems unlikely that the complete ionisation of the atom from the Migdal effect would be achievable in practice.

The fact that $p_v^\text{sum}=1$ to a high accuracy at low recoil velocities, where quadruple and higher transitions are negligible, provides a strong consistency check of our results. A further cross-check is provided by our two separate implementations, with the lower sub-panel of \cref{fig:integrated} showing the percentage difference between our two calculations ($p_v^\grasp/p_v^\bertha-1$). For ground-ground transitions the agreement is nearly perfect, and for single transitions the calculations agree within 2\%. For double transitions the agreement has greater dependence on the recoil velocity, but is never worse than 5\%.\footnote{We also compared our results for neon with Ref.~\cite{Talman:2006}, finding good agreement for $v/\alpha<0.8$. At higher recoil velocities, Ref.~\cite{Talman:2006} obtains larger single and double transition probabilities than both of our calculations (which are in close agreement).} The reason that the \grasp/\ratip\ calculation yields slightly lower probabilities is that it includes only a subset of the transitions to excited bound states\footnote{For neon, we have included orbitals up to $8s,8p,6d$; note that the excited bound orbitals from \grasp\ are also not strictly orthogonal to the continuum states from \ratip, as discussed in \cref{app:grasp}.}; \bertha, on the other hand, does not distinguish between bound and continuum states, and the basis-set set we have used is sufficiently large to achieve good convergence.

\subsection{Transition probabilities in the presence of an energy threshold}

Although the probability of multiple ionisation increases with the nuclear recoil velocity, the additional final-state electrons would only be observed by experiments with a very low electron energy threshold. The reason for this can be seen in \cref{fig:inclusive}, which shows the probabilities for double transitions in carbon (left) and argon (right) that yields either one (dashed purple line) or two (solid purple line) electrons with energies higher than $1$\,keV. The probability of obtaining two hard electrons is orders of magnitude smaller than having a single hard-electron across the whole range of nuclear recoil velocities, which extend to the maximum velocity induced by D-T neutron scattering. Accordingly, it is unlikely that Migdal events with multiple energetic electrons will be observed in either neutron-beam or DM experiments.\footnote{Ionisation by the Migdal effect necessarily leads to atomic de-excitation, which may result in additional electrons through Auger (or Coster-Kronig) emission. For light atoms, the Auger electron will also be sub-keV. See Ref.~\cite{MIGDALa} for further discussion.} 

The more relevant quantity for experiments is, therefore, the semi-inclusive probability to produce one ionisation electron above threshold, with additional bound excitations or sub-threshold ionisation electrons. This is shown by the solid grey line in \cref{fig:inclusive} with a threshold of $E_{\rm{th}}=1\,$keV for carbon (left) and argon  (right). The dashed cyan and purple curves show the contributions to the semi-inclusive probability from single and double transitions, respectively. The semi-inclusive and exclusive single ionisation curves are closely matched at low recoil velocities, but for recoil velocities $v/\alpha\gtrsim0.3$ double and higher transitions cannot be neglected and eventually dominate.

\begin{figure}[t!]
    \includegraphics[width=0.95\columnwidth]{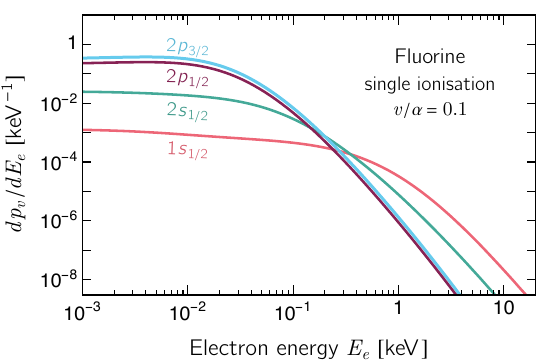}
    \par\vspace{3ex}{\par}
    \includegraphics[width=0.95\columnwidth]{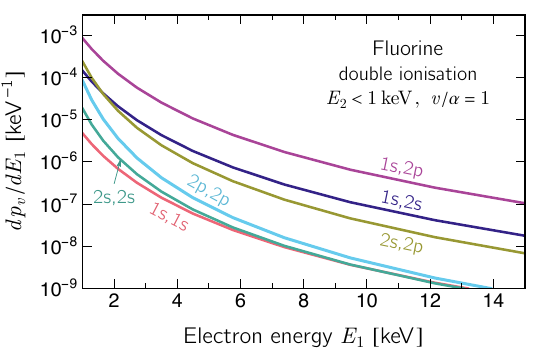}
    \par\vspace{3ex}{\par}
    \includegraphics[width=0.95\columnwidth]{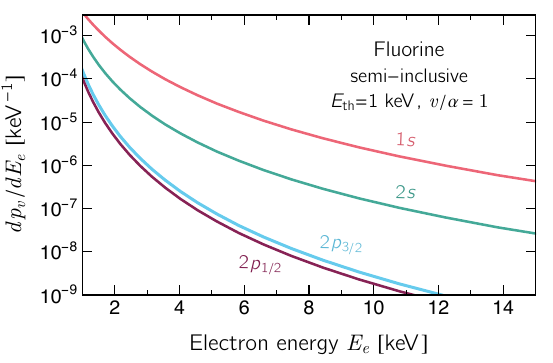}
    \caption{Contributions to the differential ionisation probability for fluorine from different initial-state orbitals. The upper panel is for single ionisation; the middle panel shows double ionisation when one ionisation electron is soft ($E_2<1~\mathrm{keV}$) while the other is hard ($E_1>1~\mathrm{keV}$); and the lower panel shows the semi-inclusive probability. The upper and lower panels label the curves by the relativistic quantum numbers while, for clarity, the middle panel uses non-relativistic notation.}
    \label{fig:fluorine-orbitals}
\end{figure}

The upper panel of \cref{fig:fluorine-orbitals} shows the contributions from individual initial-state orbitals to the single ionisation probability, as a function of the ionisation electron energy~$E_e$. Ionisation from the valence shell provides the dominant contribution at low $E_e$, while high-energy electrons are more likely to be ionised from the inner shell. Relativistic effects are small for fluorine, and the $p_{1/2}$ and $p_{3/2}$ curves are the same up to a multiplicity factor of 2/3.

The middle panel of \cref{fig:fluorine-orbitals} shows the contributions of transitions from particular pairs of initial-state orbitals to the double ionisation probability. In this panel we focus on the scenario where one ionisation electron is soft and below a threshold of $E_{\rm{th}}=1\,\mathrm{keV}$; we integrate over the energy of the soft electron and show the differential probability as a function of the hard electron's energy. For clarity, we have combined the $2p_{1/2}$ and $2p_{3/2}$ contributions and use non-relativistic notation. The combination of ionisation from the inner shell together with the valence shell ($1s,2p$) provides the dominant contribution at these energies. This behaviour is consistent with that observed for single ionisation, where ionisation from the $1s$ and $2p$ subshells dominates for hard and soft electrons, respectively. 

Finally, the lower panel of \cref{fig:fluorine-orbitals} shows the individual contributions to the sum over orbitals in \cref{eq:inclusive} for the semi-inclusive probability. Again, we see that the leading contribution is from the $1s$ matrix element. The qualitative behaviour observed in \cref{fig:fluorine-orbitals} is the same for all of the atoms we have studied, and does not change significantly with the recoil velocity up to overall normalisation.

\subsection{Validity of the dipole approximation}
\label{sec:dipole-validity}

\begin{figure}[t!]
    \includegraphics[width=0.95\columnwidth]{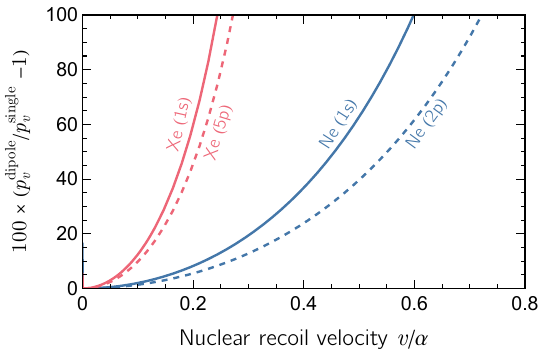}
    \par\vspace{3ex}{\par}
    \includegraphics[width=0.95\columnwidth]{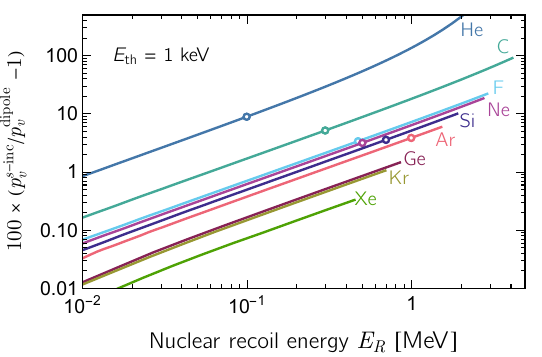}
    \caption{Upper panel: Difference between the exclusive single ionisation probabilities and the dipole approximation for the $1s$ and valence subshells of neon and xenon. Lower panel: Difference between the semi-inclusive ionisation probability and the dipole approximation for various atoms. The open circles denote where $v=\alpha$. }
    \label{fig:dipole-comparison}
\end{figure}

In \cref{fig:dipole-comparison} we provide some examples that demonstrate the accuracy of the dipole approximation. First, in the upper panel we show the difference between the exclusive single ionisation probability and the dipole approximation, $p_v^\text{dipole}/p_v^\text{single}-1$. The solid and dashed curves show ionisation from the $1s$ and valence subshells, respectively, of neon (blue) and xenon (red). As discussed in \cref{sec:dipole-approx}, and contrary to previous expectations in the literature, we see that the dipole approximation fails at roughly the same recoil velocity for both subshells. In fact, it is only the valence subshell that behaves slightly differently and all other subshells closely follow the $1s$ curve. We also see that the dipole approximation breaks down at a lower recoil velocity for xenon than neon. More generally, we find that the dipole approximation provides a good approximation to the exclusive ionisation probability only when $v\ll \alpha Z_n/n^2$, with $n$ the principal quantum number of the valence shell, which is consistent with the expectations from \cref{sec:dipole-approx}.

The lower panel of \cref{fig:dipole-comparison} shows the difference between the semi-inclusive probability and the dipole approximation, $p_v^\text{s-inc}/p_v^\text{dipole}-1$, as a function of the nuclear recoil energy for several different atoms. The endpoints of the curves correspond to the maximum nuclear recoil energy from incident D-T neutrons. Recall from \cref{sec:dipole-approx} that the dipole result is expected to provide a good approximation to the semi-inclusive probability when $v\ll \alpha Z_n$. For helium, this corresponds to $E_{R} \ll 0.4\,$MeV, which is entirely consistent with our numerical results in \cref{fig:dipole-comparison}. For heavy atoms, such as xenon, we see that the dipole approximation should be sufficiently accurate for most cases of practical interest; however, this is clearly not the case for lighter atoms, where it significantly underestimates the semi-inclusive probability for large nuclear recoil energies.

\section{Migdal Phenomenology}
\label{sec:pheno}

In this section we present two applications of our calculations. The first is to sub-GeV DM direct detection, where the nuclear recoil velocity is small, $v\ll\alpha$, and single ionisation is the dominant process. The second is to neutron scattering, where we assume the neutrons originate from D-D or D-T fusion generators and the nuclear recoil velocity can satisfy $v\simeq\alpha$, so that multiple ionisation dominates.

\subsection{Dark matter}
\label{sec:DMrates}

Consider DM that interacts with the nucleus via the usual spin-independent operator. The differential rate (per unit target mass) to produce a nuclear recoil with energy $E_R$ and an ionisation electron with energy $E_e$ factorises and can be cast in the form
\begin{equation} \label{eq:DMrate}
        \frac{d^2R}{dE_R dE_e} = \frac{\rho_\chi A^2\sigma_n}{2m_\chi\mu_{\chi n}^2} |F_{\rm{N}}|^2  \sum_{n\kappa} \frac{dp_v(n\kappa\to E_e)}{dE_e} g_\chi(v_{\rm{min}}) \,,
\end{equation}
where the local DM density is $\rho_\chi \simeq 0.3\,$GeV\,cm$^{-3}$, $m_\chi$~is the DM mass, $A$~the atomic mass number of the target, $\mu_{\chi n}$~the DM-nucleon reduced mass, $\sigma_n$~the spin-independent DM-nucleon scattering cross-section at zero-momentum transfer, and for $F_{\rm{N}}$ we use the Helm nuclear form factor~\cite{Lewin:1995rx}. The probability to ionise an electron with initial quantum numbers $(n,\kappa)$ into a final state with kinetic energy $E_e$ is $p_v(n\kappa\to E_e)$. Finally, the standard integral over the DM velocity distribution~is
\begin{equation}
    g_\chi(v_{\rm{min}}) = \int_{v_{\rm{min}}}^\infty \frac{f_\chi(\vec{v}_\chi+\vec{v}_\oplus)}{\left|\vec{v}_\chi\right|} \, d^3\vec{v}_\chi \,,
\end{equation}
where $f_\chi(\vec{v}_\chi)$ is taken to be a truncated Maxwell-Boltzmann distribution and we follow the recommendations in Ref.~\cite{Baxter:2021pqo} and set $v_0=\sqrt{2}\sigma_v=238$\,km\,s$^{-1}$ and $v_\text{escape}=544$\,km\,s$^{-1}$. We neglect the time-dependence of $\vec{v}_\oplus$, the motion of the Earth with respect to the galactic rest frame~\cite{McCabe:2013kea}. The minimum velocity of DM that can inelastically scatter to produce a nuclear recoil with energy $E_R$ and an electronic excitation of energy $E_{\rm{EM}}$ is 
\begin{equation}
    v_{\rm{min}} = \sqrt{\frac{m_{\rm{N}} E_{\rm{R}}}{2\mu^2}} + \frac{E_{\rm{EM}}}{\sqrt{2 m_{\rm{N}} E_{\rm{R}}}} \,,
\end{equation}
where $m_{\rm{N}}$ is the nucleus mass and $E_{\rm{EM}}=E_e+E_{n\kappa}$, with $E_{n\kappa}$ the (positive) binding energy of the electron before emission. Note that since $v_{\rm{min}}$ depends on $(n,\kappa)$ through~$E_{n\kappa}$, $g_\chi(v_{\rm{min}})$ should be included in the sum in \cref{eq:DMrate}.  For multiple ionisation, $E_{\rm{EM}}$ is modified to include the sum of the binding energies of each of the electrons. 

\begin{figure}[t!]
    \includegraphics[width=0.95\columnwidth]{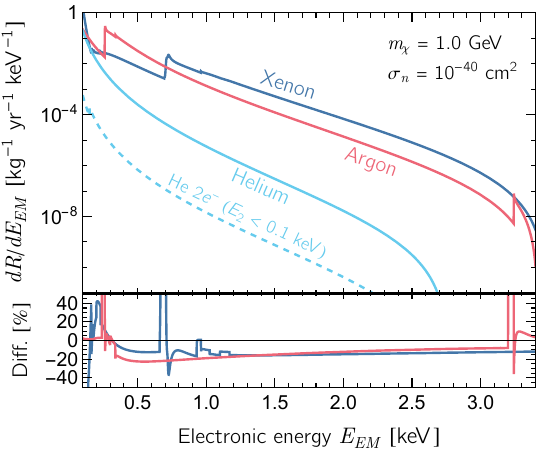}
    \caption{Differential DM scattering rate (per unit target mass) as a function of the total electronic energy for helium, argon and xenon targets, with a DM-nucleon spin-independent cross-section of $\sigma_n=10^{-40}\,$cm$^2$ and a DM mass of 1\,GeV. The lower panel shows the difference between our result and the rate obtained using the dipole approximation for $dp_v/dE_{\rm{EM}}$ from Ref.~\cite{Ibe:2017yqa}, $(dR/dE_{\rm{EM}})_\text{dipole}/(dR/dE_{\rm{EM}}) - 1$. The cyan dashed line shows the rate of double transitions in helium, where we integrated over the energy of the second, soft electron up to 0.1\,keV.}
    \label{fig:DM}
\end{figure}

In \cref{fig:DM} we show the differential DM scattering rate as a function of the total electronic energy, $E_{EM}$, for helium, argon and xenon. Results for He, relevant for the CYGNUS and NEWS-G experiments, have not been presented in the literature before. We have assumed a DM mass of $1$\,GeV and a DM-nucleon scattering cross-section of $10^{-40}\,\mathrm{cm}^2$. For argon (xenon), one can clearly identify the thresholds where ionisation from the $n=2,1$ ($n=3$) shells becomes kinematically accessible. The dashed line shows the rate of double transitions in helium, including both double ionisation and ionisation with excitation, where the second electron is soft ($E_2<0.1\,$keV). As expected, the double transition rate is highly suppressed due to the low nuclear recoil velocity ($v/\alpha \lesssim 0.1$) induced by the scattering DM. 

The dipole approximation is expected to be valid in the kinematic regime relevant for DM scattering. In the bottom sub-panel of \cref{fig:DM} we compare (for Ar and Xe) the rate obtained with our transition probabilities to the dipole approximation results of Ref.~\cite{Ibe:2017yqa}, $(dR/dE_{\rm{EM}})_\text{dipole}/(dR/dE_{\rm{EM}}) - 1$. We indeed find good agreement between the two calculations, verifying existing DM limits based on the Migdal effect. In fact, the differences are primarily due to the differing calculations of the atomic wavefunctions, rather than the use of the dipole approximation. Specifically, we use the local DHF exchange potential in contrast to the effective central potential approximation employed in Ref.~\cite{Ibe:2017yqa}. In \cref{fig:DM} we have used the theoretical values for the orbital binding energies in $E_{\rm{EM}}$, which differ between the two calculations. This is the source of the larger differences near thresholds. We provide further comparisons of our ionisation probabilities with the dipole results of Ref.~\cite{Ibe:2017yqa} in~\cref{app:compare}.

\subsection{Neutron scattering}
\label{sec:nrates}

There are several experimental proposals to test the theory underlying the Migdal effect with neutron sources. These experiments will probe an energy regime above that being exploited by DM experiments, but the systematic study of the Migdal effect in various atomic species will test theoretical predictions of the Migdal effect over a wide energy regime. We focus on the phenomenology relevant to the MIGDAL experiment since it uses the highest energy neutron sources and results in phenomenology that is most distinct from DM scattering -- and was one of the key motivations for the present work. 

\begin{figure}[t!]
    \includegraphics[width=0.95\columnwidth]{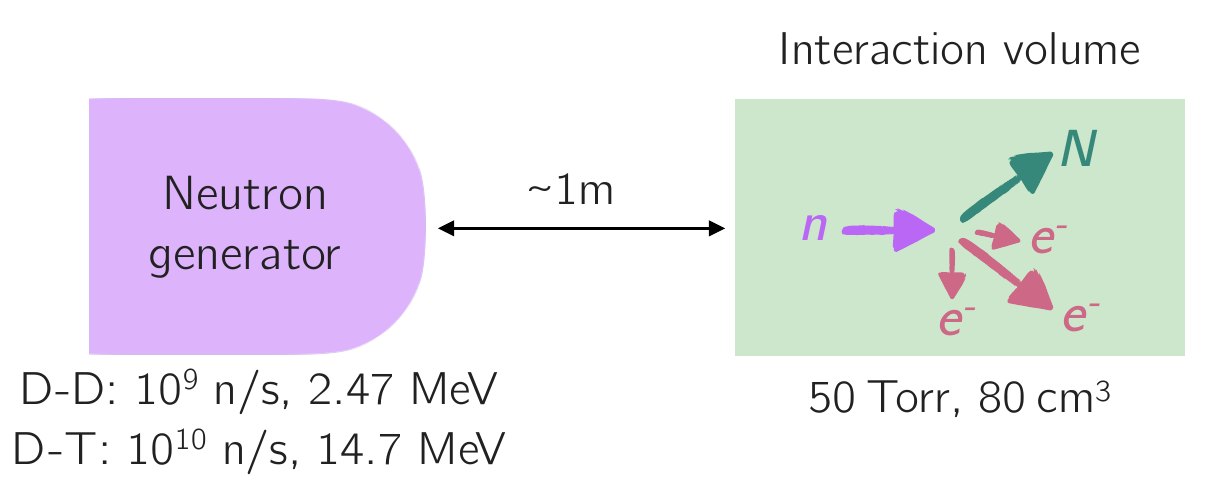}
    \caption{Schematic representation of the MIGDAL experiment. The D-D (D-T) generator emits mono-energetic neutrons at $2.47\, (14.7)$\,MeV and approximately isotropically at an intensity of $10^9\, (10^{10}$) neutrons per second. Neutrons scatter in the interaction volume contained within an OTPC, which images nuclear and electron ionisation tracks. Long and short electron arrows represent one hard ionisation electron and additional sub-threshold electrons, the dominant Migdal process for these generators.}
    \label{fig:MIGDALschematic}
\end{figure}

A schematic representation of the MIGDAL experiment is shown in \cref{fig:MIGDALschematic}. The experiment will operate with intense neutron beams from D-D and D-T fusion generators, which are directed towards an optical time-projection chamber (OTPC) situated 1~metre away for the D-T generator and 0.5~metres away for the D-D generator.\footnote{We ignore the mild dependence of the neutron intensity on the orientation of the generator and assume that the neutrons are emitted isotropically. We use the neutron energies for the generator orientations that will be used by the MIGDAL experiment.} The OTPC will be filled with low-pressure gas: initially pure CF$_4$ and later CF$_4$-based mixtures with noble elements or Si- or Ge-compounds. Nuclear and electron recoils within the low-pressure gas result in ionisation tracks. The tell-tale sign of a Migdal event is a nuclear recoil track and an electron ionisation track emerging from a common vertex. The ionisation tracks must be sufficiently long to discriminate between electron and nuclear recoils; at a pressure of 50~Torr, this implies an electron energy threshold of approximately 5\,keV and a nuclear recoil threshold of around 150\,keV~\cite{MIGDALa}.

The rate for neutron-induced Migdal events in a gas-mixture again factorises and can be expressed as
\begin{equation}
    \frac{d^2R}{d E_R dE_e} =\phi_n\, \sum_{i} N^i_T\, \frac{d \sigma^i_s}{d E_{R}} \sum_{n\kappa} \frac{dp^i_v(n\kappa\to E_e)}{dE_e} \,,
\end{equation}
where $\phi_n$ is the neutron flux, the sum over $i$ runs over all species in the gas-mixture (e.g.\ $i=\{\mathrm{C},\mathrm{F},\mathrm{Ar}\}$ in a CF$_4$+Ar mixture), and $N^i_T$, $\sigma^i_s$, and $p^i_v(n\kappa\to E_e)$ are the number of target atoms in the interaction volume, the neutron-nucleus cross-section, and the transition probability for the $i$-th atomic species, respectively. The number of target atoms is calculated assuming the gas is at ambient temperature (293.15\,K) and the interaction volume is 80\,cm$^3$ so that, for instance, CF$_4$ gas at 50~Torr contains $1.3\times10^{20}$ molecules. To calculate $N_T^i$, we treat each CF$_4$ molecule as one carbon atom and four fluorine atoms.

An important difference with respect to DM scattering is that the neutron has sufficient energy to excite the nucleus. Immediately after the scattering, the excited nucleus is moving with respect to the electron-cloud and so can still lead to electron emission through the Migdal effect. Therefore, $\sigma_s$ should include the contributions from all reactions that result in the topology in \cref{fig:MIGDALschematic}: elastic scattering, inelastic scattering, $(n,2n)$ reactions and radiative capture reactions, since all produce bare nuclear recoils (i.e.\ without accompanying charged tracks) and the photon released during de-excitation of the nucleus escapes the low-pressure gas without interacting. Differential and integrated neutron-nucleus cross-sections that we use in this work are given in \cref{sec:nxsecs}.

Treating the molecule as a sum of discrete atomic nuclei should be a very accurate approximation in the context of the neutron-nucleus interaction, since the de Broglie wavelength of the neutron at D-D or D-T energies is $\sim 10^{-14}\,\mathrm{m}$, which is orders of magnitude smaller than the C-F bond length. The approximation of using the atomic result for the transition probability, $p^i_v(n\kappa\to E_e)$, rather than performing a molecular calculation, requires a more careful justification. As a starting point, it has been shown that the general form of the transition matrix element, \cref{eq:migdal_formula}, remains the same for both atoms and molecules (up to corrections of order $m_e/m_{\mathrm{N}}$)~\cite{Lovesey:1982}. It is also common practice to model the molecular electronic wavefunctions in terms of anti-symmetric products of single-electron wavefunctions (molecular orbitals). Taken together, this implies that the formalism in \cref{sec:derivation} extends to the case of molecules: this particularly applies to the derivation of the semi-inclusive rate where the only assumption made about the soft-electron wavefunctions is that they are orthogonal and complete, which holds for both atoms and molecules.

In the context of the MIGDAL experiment, we expect that the atomic result for the semi-inclusive transition probability will provide a good approximation to the molecular result. This is because the MIGDAL experiment employs an $\mathcal{O} (\mathrm{keV})$ energy threshold, so as \cref{fig:fluorine-orbitals} demonstrates, the transition probability is dominated by the core (inner-most) electrons. This is important because, after molecular bonding, the core electrons in the CF$_4$ molecule are only slightly modified from their atomic forms. This is evidenced by the experimental values of the binding energies: the $1s$ binding energy in the fluorine atom is 697 eV~\cite{lbl_xray_booklet}, while the $1t_2$/$1a_1$ states in CF$_4$ (the core states equivalent to $1s$) have a binding energy of 695 eV~\cite{jolly1984core}. We therefore expect that when the core electrons dominate the scattering rate, which is the case for the MIGDAL experiment, the atomic transition probability will provide a reasonable approximation to the full molecular result. Further work is warranted to quantify the level of agreement but such a study lies beyond the scope of this work.

\begin{figure}[t!]
    \includegraphics[width=0.95\columnwidth]{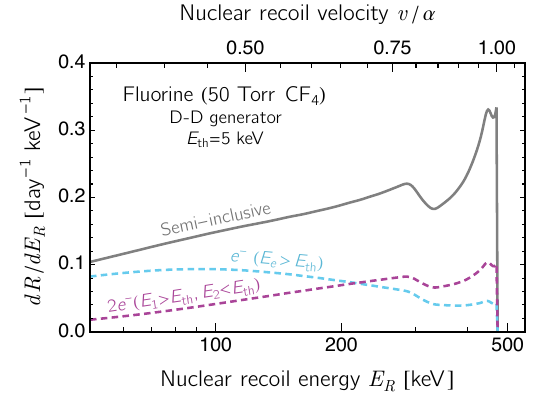}
     \par\vspace{3ex}{\par}
    \includegraphics[width=0.95\columnwidth]{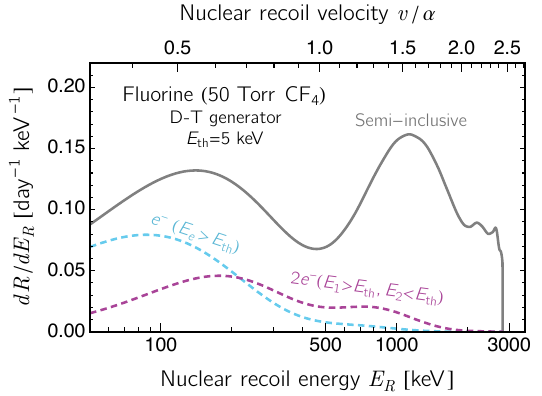}
    \caption{Differential scattering rate as a function of the nuclear recoil energy for fluorine in CF$_4$ gas at 50~Torr from a D-D (top) and D-T (bottom) neutron generator.  Electron energies above the threshold of $E_{\rm{th}}=5\,\mathrm{keV}$ have been integrated over. The solid grey curve shows the semi-inclusive rate while the dashed cyan and purple curves show the contributions from single and double transitions, respectively.}
    \label{fig:neutron-rates}
\end{figure}

With the above assumptions, the result for the fluorine differential scattering rate with an electron energy threshold of $E_{\rm{th}}=5\,\mathrm{keV}$ in CF$_4$ gas at 50~Torr is shown in the upper (lower) panel of \cref{fig:neutron-rates} for the D-D (D-T) neutron generator. The dotted lines show the rate for single ionisation (cyan) and double transitions (purple), while the solid grey line shows the semi-inclusive rate. In both cases the semi-inclusive curve matches the sum of the single and double rates at $v/\alpha\approx0.3$. However for higher values of $v/\alpha$, and most dramatically in the case of the D-T generator, the semi-inclusive rate departs significantly from the combined single and double rates. This behaviour is consistent with that shown in \cref{fig:inclusive} for the transition probabilities.

\begin{figure}[t!]
    \includegraphics[width=0.95\columnwidth]{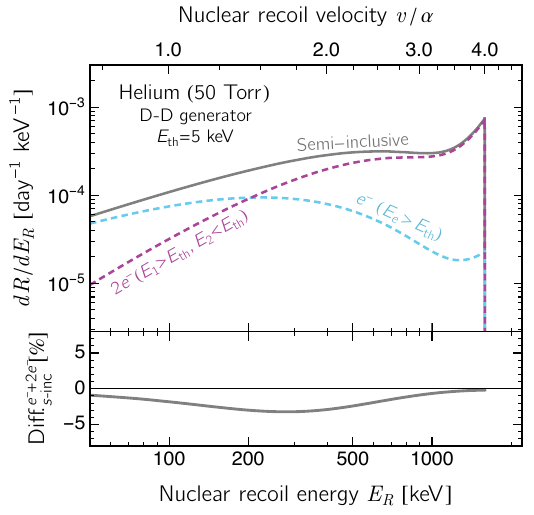}
    \caption{Differential scattering rate as a function of the nuclear recoil energy for helium gas at 50~Torr from a D-D neutron generator. Electron energies above $E_{\rm{th}}=5\,\mathrm{keV}$ have been integrated over. The solid grey curve shows the semi-inclusive rate while the dashed cyan and purple curves show the contributions from single and double transitions, respectively. The lower panel shows the percentage difference between the sum of the exclusive single and double transition rates and the semi-inclusive rate, $(dR/dE_R)_{e^-+2e^-}/(dR/dE_R)_{\text{s.-inc}}-1$.}
    \label{fig:Heneutron-rates}
\end{figure}

Helium provides an interesting example since we can directly compare the semi-inclusive calculation with the sum of the exclusive transitions. The differential scattering rate induced by a D-D generator directed at helium gas at 50~Torr is shown in \cref{fig:Heneutron-rates}.\footnote{Experimentally, it may be more favourable to operate with helium gas at higher pressure~\cite{MIGDALa}, but for ease of comparison with other elements, we present results for 50~Torr.} The solid grey line again shows the semi-inclusive rate, while the dashed purple line shows the double transition rate. As expected, the single ionisation rate dominates at low values of $v/\alpha$, while the double transition rate dominates at larger values. The lower panel shows the percentage difference between the sum of exclusive transitions and the semi-inclusive rate, $(dR/dE_R)_{e^-+2e^-}/(dR/dE_R)_{\text{s.-inc}}-1$. We find good agreement at the level of a few \% or better across the whole nuclear recoil range; $(dR/dE_R)_{e^-+2e^-}$ slightly underestimates $dR/dE_R|_{\text{s.-inc}}$, with the small difference arising because the excited bound orbitals from \grasp\ and the continuum states from \ratip\ are not strictly orthogonal, as discussed in \cref{app:grasp}, so the ionisation with excitation rate is slightly underestimated.

\begin{figure}[t!]
\par\vspace{3ex}{\par}
    \includegraphics[width=0.95\columnwidth]{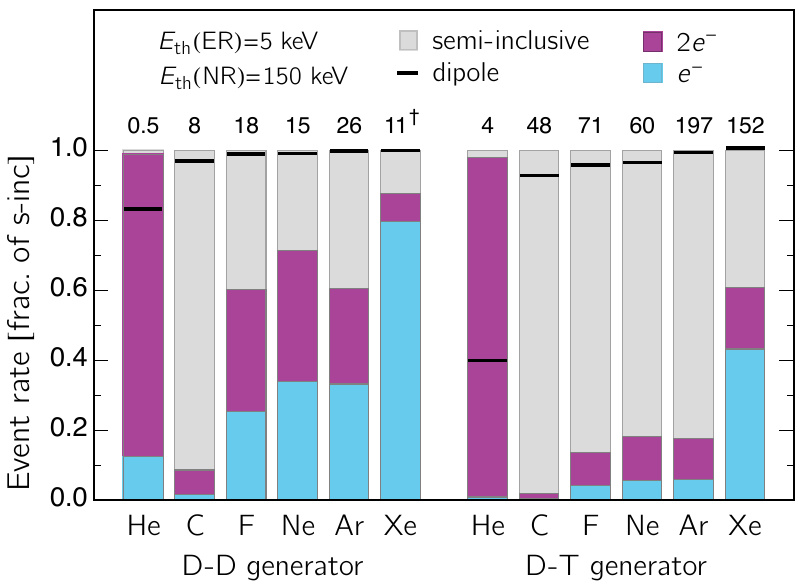}
    \caption{Event rates for several atomic elements with D-D (left set of columns) and D-T (right set of columns) generators and electron and nuclear recoil energy thresholds of 5\,keV and 150\,keV, respectively. The numbers above each column give the events-per-day for 50~Torr gas. The coloured columns show the stacked event rates for single (cyan) and double (purple) transitions relative to the semi-inclusive rate. The black line in each column shows the dipole approximation event rate relative to the semi-inclusive rate. \newline$^\dagger$ For xenon with the D-D generator $E_\mathrm{th}(\mathrm{NR})=50$\,keV.}
    \label{fig:Int-neutron-rates}
\end{figure}

In \cref{fig:Int-neutron-rates}, we compare the number of Migdal events induced by a D-D generator (left set of columns) and a D-T generator (right set of columns) in several single-species gas targets. The expected number of events-per-day is listed above the column for each element, assuming a gas pressure of 50~Torr and an experimental set-up as in \cref{fig:MIGDALschematic}. These were obtained by integrating the semi-inclusive rate above an electron energy threshold of 5\,keV and a nuclear recoil energy threshold of 150\,keV. The only exception is for xenon with the D-D generator where the end-point energy is approximately 75\,keV, so we instead integrated over nuclear recoil energies above 50\,keV. With both neutron generators, we see that there is the possibility of inducing tens or hundreds of Migdal events over a short data-taking period consisting of a few days.

The cyan columns in \cref{fig:Int-neutron-rates} show the single ionisation event-rate relative to the semi-inclusive rate. The purple columns show the event-rate from double transitions\footnote{For all elements except xenon, this includes both double ionisation and ionisation with excitation, with the latter contributing $\sim5\%$ to the double transition rate; only double ionisation is included for xenon.}, again relative to the semi-inclusive rate, stacked above the single ionisation contribution.  For helium the single plus double transition rate (i.e.\ the total height of the cyan and purple columns) almost exactly matches the semi-inclusive rate, consistent with \cref{fig:Heneutron-rates}. However, for all of the other gases it underestimates the semi-inclusive rate. The discrepancy is largest for lighter atoms (helium excluded) and for the D-T generator where the nuclear recoils extend to larger values of~$v/\alpha$.

Finally, the black line in each column of \cref{fig:Int-neutron-rates} indicates the event-rate obtained using the dipole approximation, relative to the semi-inclusive rate. The general trend is consistent with that shown in \cref{fig:dipole-comparison}: the dipole approximation provides a good approximation for the heavier atoms, while for lighter atoms it significantly underestimates the semi-inclusive rate, especially for the larger nuclear recoil energies induced by the D-T generator.

\section{Conclusions}
\label{sec:conc}

The excitation or emission of an electron after a sudden jolt to the atomic nucleus by an electrically-neutral projectile is known as the Migdal effect. In recent years, the Migdal effect has gained prominence in direct detection searches for sub-GeV DM, and there are now several proposals to measure the effect using neutron sources. The parameter that characterises the probability of electron emission is~$v/\alpha$, the dimensionless ratio of the nuclear recoil velocity relative to the fine-structure constant multiplied by the speed-of-light. Previous studies have focused on the regime where $v/\alpha\ll1$, which is most relevant for DM direct detection searches. We have advanced the theory underlying the Migdal effect in the $v\simeq\alpha$ regime, introducing the semi-inclusive probability that is essential for accurate predictions in fast neutron scattering.

We have undertaken two independent calculations of the Migdal effect, solving the Dirac-Hartree-Fock equations using the finite difference and basis-set methods as implemented in \grasp\ and \bertha, respectively. We have calculated electron-transition probabilities for noble elements from helium to xenon, as well as for carbon, silicon, germanium and fluorine. In addition to calculating single ionisation probabilities, which are in good agreement with previous calculations when $v/\alpha\ll1$, we have undertaken calculations of multiple ionisation. We have shown that multiple ionisation dominates when $v\sim\alpha$, but that the probability of obtaining multiple electrons with energies above an $\mathcal{O}(\mathrm{keV})$ experimental threshold is extremely small. Accordingly, we have emphasised the importance of the \textit{semi-inclusive} transition probability, which accounts for all processes that yield one hard electron, with any number of additional soft, sub-threshold electrons. We have also clarified the role of the dipole approximation. While this approximation is only formally defined for single transitions, we have found that it, perhaps surprisingly, yields a good estimate of the semi-inclusive probability for all but the lightest atoms. 

We have applied our results for the Migdal effect in the context of both DM direct detection and neutron scattering experiments. In the DM case, we found good agreement between our single ionisation calculations and previous results. Furthermore, we confirmed that the double ionisation rate is highly suppressed, owing to the low nuclear recoil velocities involved, and can be safely ignored. We have performed the first calculations of the Migdal effect in DM-helium scattering, which will enable experiments that use this element to increase their sensitivity to sub-GeV DM. For neutron scattering, we focused on the phenomenology relevant to the MIGDAL experiment, which employs D-D and D-T neutron generators. At the nuclear recoil velocities induced by these generators, we have shown that the multiple ionisation rate is significant and cannot be ignored: it is imperative that the semi-inclusive probabilities are used to provide the most accurate description of the Migdal effect.

Our work has been carried out in the context of atomic systems. While we have argued that the atomic calculations should provide a good approximation for the core-electrons in $\mathrm{CF}_4$, a molecular gas that will be used in the MIGDAL experiment, work remains to extend the formalism so that it applies to all electrons in the molecular system. Similarly, we have ignored complications due to the electronic band structure in liquid noble elements, relevant for DM direct detection experiments. Further work is warranted to extend the theory to these systems. We hope to address some of these issues in the future.


\noindent\\
\textbf{Data Access Statement:}
The data supporting the findings reported in this paper are openly available from the GitHub and Zenodo repositories at~\cite{Git_results}.


\begin{acknowledgments}
PC is supported by the Australian Research Council Discovery Early Career Researcher Award DE210100446. MJD is supported by the Australian Research Council. This work was supported by the Australian Research Council through the ARC Centre of Excellence for Dark Matter Particle Physics, CE200100008. CM is supported by the UKRI’s Science and Technology Facilities Council (awards ST/N004663/1, ST/V001876/1, ST/T000759/1). We are grateful to members of the MIGDAL Collaboration for discussions and particularly thank T.~Marley (Imperial) for generating the neutron scattering cross-sections used in this work. For the purpose of open access, the authors have applied a Creative Commons Attribution (CC BY) licence to any Author Accepted Manuscript version arising from this submission.
\end{acknowledgments}


\appendix

\section{Atomic Physics}
\label{app:atomic}

\subsection{Relativistic single-electron matrix elements}
\label{app:relativistic-MEs}

To evaluate the Migdal transition matrix element, we require single-electron matrix elements of the form
\begin{equation}
    M_{n \kappa m}^{n' \kappa' m'} \equiv \langle n',\kappa', m'|\exp(im_e\bm{v}\cdot\bm{r})|n,\kappa, m\rangle \,, 
\end{equation}
where $\langle \bm{r}|n,\kappa,m \rangle$ is a four-component atomic Dirac spinor that satisfies
\begin{align}
    \langle \bm{r}|h_D|n,\kappa,m \rangle &= \langle \bm{r}|{\bm\alpha}\cdot \bm{p} + \beta m_e + V(\bm{r})|n,\kappa,m \rangle \nonumber \\
    &= E_{n}\langle \bm{r}|n,\kappa,m \rangle \,,
\end{align} 
with $n$ denoting the principal quantum number, $m$ the $J_z$ quantum number, $V(\bm{r})$ the potential, and $\bm\alpha$ and $\beta$ are the $4\times4$ Dirac matrices~\cite{Dirac:1928hu}.  The quantum number $\kappa$ is the eigenvalue of the operator $K=-1-\bm{\sigma}\cdot \bm{l}$ and can be written as
\begin{equation}
    \kappa =
    \begin{cases}
     -(l+1) & \, j=l+1/2 \quad (\kappa <0) \\
     +l &  \, j=l-1/2 \quad (\kappa>0) \,,
    \end{cases}
\end{equation}
with $l$ and $j$ the orbital and total angular momentum quantum numbers respectively. For continuum states the discrete label $n$ is replaced with the continuous label $E$. We normalise our continuum spinors with respect to energy, $\int dE \braket{\psi_{E',\kappa',m'}|\psi_{E,\kappa,m}}=\delta(E-E')$. 

The four-component spinors are separable in radial and spin-angular coordinates and can be written as (see e.g.~\cite{GrantBook})
\begin{equation}
    \langle \bm{r}| n,\kappa,m\rangle = \frac{1}{r}
    \left(
    \begin{array}{l}
        P_{n,\kappa}(r) \chi_{\kappa,m}(\vartheta,\varphi) \\
        i Q_{n,\kappa}(r) \chi_{-\kappa,m}(\vartheta,\varphi)
    \end{array}
    \right) \,,
\end{equation}
where $P_{n,\kappa}(r)$ and $Q_{n,\kappa}(r)$ are, respectively, the large and small component radial functions and $\chi_{\pm\kappa,m}(\vartheta, \varphi)$ are two-component spin-angular functions.

As usual, the operator $\exp(im_e\bm{v}\cdot\bm{r})$ is written as the spherical tensor expansion
\begin{equation} \label{eq:spherical-tensor-expansion}
    \exp(im_e\bm{v}\cdot\bm{r}) = 4\pi \sum_{L,M} i^L j_L(m_evr)Y_L^{M*}(\hat{\bm{v}})Y_L^{M}(\hat{\bm{r}}) \,,
\end{equation}
where $j_L(x)$ is the spherical Bessel function. The spin-angular parts of the matrix element are independent of $\mbox{sgn}(\kappa)$ and $\mbox{sgn}(\kappa')$, which allows us to write the matrix element in the form
\begin{multline} \label{eq:1e-ME}
    M_{n \kappa m}^{n' \kappa' m'} = \sqrt{4\pi}\sum_{L,M}i^L \sqrt{2L+1} \, Y_L^M(\hat{\bm{v}}) \, d^L_M(j',m';j,m) \\
    \times \int_0^{\infty} dr\, j_L(m_evr) \big[ P_{n,\kappa}(r)P_{n',\kappa'}(r) \\
    \qquad + Q_{n,\kappa}(r)Q_{n',\kappa'}(r) \big] \,.
\end{multline}
The angular coefficients can be determined via the Wigner-Eckart theorem and are given by
\begin{multline} \label{eq:angular-coeffs}
    d^L_M(j',m';j,m) = (-1)^{2j'-m'+1/2} \, [j,j']^{1/2} \, \Pi^e(\kappa,\kappa'; L) \\
    \times\left(
    \begin{array}{rrr}
        j' & L & j \\
        -m' & M & m
    \end{array}
    \right)
    \left(
    \begin{array}{rrr}
        j' & L & j \\
        1/2 & 0 & -1/2
    \end{array}
    \right) \,,
\end{multline}
where $[j,j']=(2j+1)(2j'+1)$, and $\Pi^e(\kappa,\kappa';L)$ implements the even-parity selection rule and is equal to one if $l+l'+L$ is even and zero otherwise. In practice, the limits on the sum over $L$ in \cref{eq:1e-ME} are determined by the selection rules of the $3j$-symbols that appear in \cref{eq:angular-coeffs}.

\subsection{Transition probabilities with configuration state functions}
\label{app:open-shell}

As usual, we describe atomic wavefunctions with configuration state functions. These are linear combinations of Slater determinants that, in the relativistic case, are eigenfunctions of energy, the total angular momentum operators, $\bm{J}$ and $J_z$, and parity. In this appendix we generalise the results of \cref{sec:derivation}, which were derived for wavefunctions consisting of a single Slater determinant. We write the initial and final-state atomic wavefunctions as
\begin{align} \label{eq:CSF}
    \big|\Psi_i\,\big\rangle &= \sum_\gamma C_i^\gamma \big|\Psi_i^\gamma\big\rangle \,, \notag \\
    \big|\Psi_f\big\rangle &= \sum_\gamma C_f^\gamma \big|\Psi_f^\gamma\big\rangle \,,
\end{align}
with ${\ket{\Psi_i^\gamma}=\ket{\psi_{a(\gamma)_1} \ldots \psi_{a(\gamma)_N}}}$, ${\ket{\Psi_f^\gamma}=\ket{\chi_{b(\gamma)_1} \ldots \chi_{b(\gamma)_N}}}$ single-Slater-determinant wavefunctions and $C_i^\gamma$, $C_f^\gamma$ constant coefficients.

\subsubsection{Exclusive transition probability}

Using the above expressions for the initial and final state wavefunctions, the generalisation of the exclusive transition probability from \cref{eq:probDet} is
\begin{equation}
    p_v\left(\ket{\Psi_i} \to \ket{\Psi_f} \right) = \bigg|\sum_{\gamma,\gamma'} (C_f^{\gamma'})^* C_i^\gamma \det\big(M^{\gamma'\gamma}\big)\bigg|^2 \,.
\end{equation}
As in \cref{eq:migdal_matrix}, $M^{\gamma'\gamma}$ is an $N \times N$ matrix of single-electron matrix elements 
\begin{equation}
    \left(M^{\gamma'\gamma}\right)_{\beta\alpha} = \big\langle\chi_{b(\gamma')_\beta}\big| \exp(im_e\bm{v}\cdot\bm{r}) \big|\psi_{a(\gamma)_\alpha}\big\rangle \,,
\end{equation}
where $\ket{\psi_{a(\gamma)_\alpha}}$ and $\ket{\chi_{b(\gamma')_\beta}}$ are occupied orbitals in $\ket{\Psi_i^\gamma}$ and $\ket{\Psi_f^{\gamma'}}$, respectively.

\subsubsection{Semi-inclusive transition probability}

In \cref{sec:derivation} we defined the semi-inclusive transition probability, which includes all final states containing a single continuum electron with $E_e>E_\mathrm{th}$ and any number of additional sub-threshold excitations. To perform the sum over final states it is more convenient to work in the determinant basis for the final states. The initial-state atomic wavefunction, on the other hand, is given by a particular linear combination of determinants, as in \cref{eq:CSF}. The semi-inclusive probability can then be written as
\begin{widetext}
\begin{align} \label{eq:inclusive-CSF}
    p_v(\ket{\Psi_i} \to \ket{\chi_{b_1} X_\text{soft}}) &= \frac{1}{(N-1)!}\sum_{b_2, \ldots, b_N}^{(E<E_\text{th})} \bigg| \Big\langle\chi_{b_1}\ldots\chi_{b_N}\Big| e^{im_e\bm{v}\cdot\sum_k \bm{r}_k} \Big|\Psi_i\Big\rangle \bigg|^2 \notag \\
    &\approx \Big\langle\Psi_i\Big| e^{-im_e\bm{v}\cdot\sum_k \bm{r}_k} \bigg( \frac{1}{(N-1)!} \sum_{b_2, \ldots, b_N}^\text{all states} \Big|\chi_{b_1}\ldots\chi_{b_N} \Big\rangle \Big\langle \chi_{b_1}\ldots\chi_{b_N}\Big| \bigg) e^{im_e\bm{v}\cdot\sum_k \bm{r}_k} \Big|\Psi_i\Big\rangle \notag \\
    &= \sum_{\gamma,\gamma'} \Big( (C_i^{\gamma'})^* C_i^\gamma \,\big\langle\Psi_i^{\gamma'}\big| e^{-im_e\bm{v}\cdot\bm{r}_1} \big|\chi_{b_1}\big\rangle \mathds{1}_{(N-1)} \big\langle\chi_{b_1}\big| e^{im_e\bm{v}\cdot\bm{a_1}} \big|\Psi_i^\gamma\big\rangle \Big) \,.
\end{align}
\end{widetext}
In the second line we have approximated $\sum_{b_2,\ldots,b_N}^{(E<E_{\mathrm{th}})} \to \sum_{b_2,\ldots,b_N}^\text{all states}$, as discussed in \cref{sec:derivation}. In going from the second to the third line we have substituted \cref{eq:CSF} for the initial-state wavefunction and used the completeness of the $\{\ket{\chi_b}\}$. The identity operator that acts on the subspace of the remaining $N-1$ electrons is denoted by $\mathds{1}_{(N-1)}$. The above expression for the semi-inclusive probability can be further simplified in specific instances. One such case is when each of the determinants that make up the initial-state wavefunction differ by two or more orbitals. This occurs for the group 14 (group IV) elements we consider (C, Si, Ge), where the atomic ground state has $J=0$ and in relativistic $jj$-coupling is a linear combination of valence configurations $(np_{1/2})^2$ and $(np_{3/2})^2$. In this case, the cross-terms in \cref{eq:inclusive-CSF} vanish and the semi-inclusive probability is 
\begin{multline}
    p_v(\ket{\Psi_i} \to \ket{\chi_{b_1} X_\text{soft}}) \approx \\
    \sum_\gamma \big|C_i^\gamma\big|^2 \sum_{\alpha=1}^N \Big| \big\langle\chi_{b_1}\big| e^{im_e\bm{v}\cdot\bm{r}} \big|\psi_{a_\alpha}\big\rangle \Big|^2 \,.
\end{multline}
Finally, for the group 17 and 18 (group VII and VIII) elements the ground state wavefunction is described by a single Slater determinant and the semi-inclusive probability is simply given by \cref{eq:inclusive}.

\subsection{Implementation in \grasp\ \& \ratip}
\label{app:grasp}

This appendix briefly describes our computation of the atomic wavefunctions using the \grasp~\cite{Jonsson:2007, Jonsson:2013, Fischer:2019} and \ratip~\cite{Fritzsche:2012} packages.

\grasp\ is an implementation of the multi-configuration Dirac-Hartree-Fock method in which wavefunctions are expressed as a weighted sum of CSFs. The integro-differential DHF equations for the radial functions are solved using finite-difference methods as part of an iterative, self-consistent field procedure. \grasp\ employs an exponential radial grid; we use the default value for the first grid point, $r_0=2.0\times10^{-6}/Z$ with $Z$ the atomic number, and a step size in the range $h=0.006-0.008$ depending on the element. The nuclear charge distribution is modelled by the default Fermi distribution. The Breit interaction and vacuum polarisation and self-energy corrections are included in the configuration interaction.

We initially solve for the atomic ground state wavefunction in an optimal level calculation. The radial functions for the excited orbitals are then obtained via an extended optimal level calculation that includes configurations with a single excitation from the valence subshell. This is repeated several times, successively increasing the maximum principal quantum number, $n$, with all radial functions with lower $n$ held fixed. The resulting set of radial functions is then used to construct all excited and ionised states. We therefore do not allow for relaxation of the orbitals; this is consistent with the sudden approximation employed in the derivation of the Migdal matrix element. 

For ionised states, the radial functions for the continuum spinors are calculated using \ratip, which is also based on the \grasp\ implementation of the multi-configuration DHF method. \ratip\ employs a radial grid that is exponential at small $r$ and transitions to linear at large $r$. We fix the grid parameters to be $r_0=2.0\times10^{-6}/Z$, $h=0.006$, and $h_p=0.0033$. For our maximum electron energy of $E_e=20$\,keV this corresponds to 50 grid points per wavelength at large $r$. The continuum radial functions are each obtained in an optimal level calculation using a configuration with a single excitation from the valence subshell. This set of continuum radial functions is then used to describe all ionised states. 

\ratip\ enforces orthogonality between the continuum wavefunctions and the occupied ground state orbitals. However, the excited bound orbitals obtained using \grasp\ and the continuum orbitals from \ratip\ are not strictly orthogonal. As a result, the basis of single-electron wavefunctions obtained using \grasp\ and \ratip\ does not approach completeness. This does not impact the single, double, and semi-inclusive ionisation probabilities we are primarily interested in, but does affect the total integrated probabilities (except for helium where the integrated probabilities can be expressed purely in terms of ground-state matrix elements). This is one of the reasons that the alternative basis-set approach of \bertha\ provides a valuable cross-check of our results.

To evaluate the transition matrix elements in \cref{sec:derivation,app:open-shell}, the wavefunctions need to be converted from the CSF basis to the determinant basis, for which we use the {\tt CESD} component of \ratip. Finally, our transition probabilities include contributions up to $L=10$ ($L=4$ for xenon double ionisation), which we find is sufficient to achieve good convergence up to the maximum recoil velocity for D-T neutrons. The one exception to this is helium, for which we include up to $L=30$. This results in excellent convergence up to $v/\alpha\sim6$, while from $v/\alpha=6$ to the D-T endpoint at $v/\alpha\simeq9.6$ conservation of probability is still maintained at the level of 4\% or better.

\subsection{Implementation in \bertha}
\label{app:bertha}

In this appendix we briefly describe the use of basis-set methods to compute the atomic wavefunctions and single-electron matrix elements, as implemented in the software package \bertha~\cite{Quiney:1998}.

The radial functions are expanded in a finite basis set. A conventional choice is the so-called $G$-spinor basis set, which takes the form
\begin{align}
    f^\mathrm{L}_{i,\kappa}(r) &= N^\mathrm{L}_{i,\kappa} r^{\ell+1}\exp(-\lambda_{i,\kappa}r^2) \,, \\
    f^\mathrm{S}_{i,\kappa}(r) &= N^\mathrm{S}_{i,\kappa}\left(\frac{d}{dr}+\frac{\kappa}{r}\right)f^\mathrm{L}_{i,\kappa}(r) \label{kinetic_balance} \,, \\
    &= N^{\mathrm{S}'}_{i,\kappa}\left((\ell+\kappa+1)-2\lambda_{i,\kappa}r^2\right)r^{\ell}\exp(-\lambda_{i,\kappa}r^2) \nonumber \,,
\end{align}
where $N^\mathrm{L}_{i,\kappa}$, $N^\mathrm{S}_{i,\kappa}$ and $N^{\mathrm{S}'}_{i,\kappa}$ are normalisation constants that prove useful in maintaining numerical accuracy in the calculation of matrix elements. The basis set parameters, $\{ \lambda_{i,\kappa}\}$, are chosen following well-established practices that generate accurate bound-state energies and which approach completeness in a systematic fashion. 

In this basis, the radial amplitudes are expressed in terms of the basis set by
\begin{align}
    P_{n,\kappa}(r) &= \sum_{i=1}^{N_{\kappa}}c^\mathrm{L}_{n,\kappa,i} \, f^\mathrm{L}_{i,\kappa}(r) \,, \\
    Q_{n,\kappa}(r) &= \sum_{i=1}^{N_{\kappa}}c^\mathrm{S}_{n,\kappa,i} \, f^\mathrm{S}_{i,\kappa}(r) \,,
\end{align}
where $N_{\kappa}$ is the rank of the expansion and $2N_{\kappa}$ is the dimension of the matrix representation of the Dirac operator for symmetry-type $\kappa$. The expansion coefficients are determined by the solution of a generalised matrix eigenvalue equation of the general form
\begin{equation} \label{eq:MEigen}
    \bm{F}_{\kappa}\bm{c}_{n,\kappa}= E_{n,\kappa}\bm{S}_{\bm{\kappa}}\bm{c}_{n,\kappa} \,. 
\end{equation}
The matrix $\bm{F}_{\kappa}$ is a matrix representation of the Dirac-Hartree-Fock operator, and includes the effects of the Coulomb interactions in the self-consistent field. The matrix $\bm{S}_{\kappa}$ is the block-diagonal Gram (or overlap) matrix in the given $G$-spinor basis set.   

With this basis set, the single-electron matrix elements in \cref{eq:1e-ME} can be expressed in closed form in terms of a single class of radial integral,
\begin{align} \label{eq:ilm} 
    I(L,M;\zeta) &= \int_0^{\infty} r^{M}j_L(m_evr)\exp(-\zeta r^2) dr \,, \nonumber \\
    &= \frac{(m_e v)^L}{2^{L+2}}\frac{\sqrt{\pi}}{\zeta^{(M+L+1)/2}}\frac{\Gamma(\frac{M+L+1}{2})} {\Gamma(L+\frac{3}{2})} e^{-\frac{1}{4\zeta}(m_e v)^2} \notag \\
    & \times \mathcal{M}\left[\frac{L-M}{2}+1,L+\frac{3}{2}; \frac{(m_e v)^2}{4\zeta}\right] \,,
\end{align}
where $\mathcal{M}[a,b;x]$ is the confluent hypergeometric function and $\zeta$ is a real, positive parameter derived from the constituent basis set parameters $\lambda_{i\kappa}$ and $\lambda_{j\kappa'}$. The parameters $L$ and $M$ are both odd or even positive integers and satisfy the subsidiary condition that $M-L$ is an even integer greater than 2, so that the confluent hypergeometric function always takes the form of a polynomial with a finite number of terms. The radial matrix elements are evaluated by taking appropriate linear combinations of the primitive integrals in \cref{eq:ilm}.

The solution of the DHF equations using $G$-spinors is achieved using the computer program \bertha. For a basis set of rank $N_{\kappa}$, the solution of equations of the form \eqref{eq:MEigen} generates a set of $N_{\kappa}$ positive-energy states and $N_{\kappa}$ negative-energy states for symmetry-type $\kappa$. The positive-energy states can be further categorised as being bound-states if $E_{n,\kappa}< mc^2$, or virtual states if $E_{n,\kappa}> mc^2$.

\section{\texorpdfstring{Comparison with previous calculations}{Comparison with previous calculations}}
\label{app:compare}

\begin{figure*}[t!]
    \includegraphics[width=0.63\columnwidth]{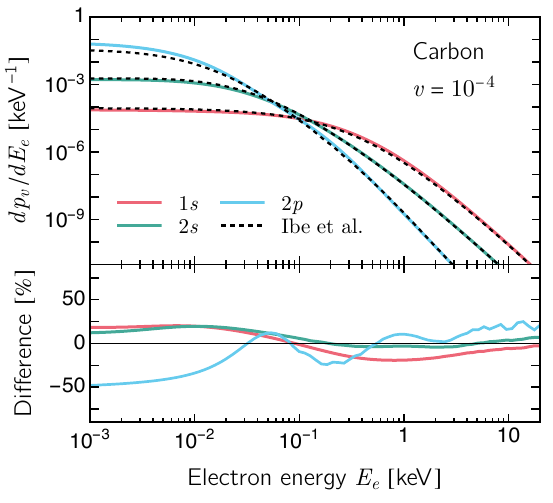}
    \includegraphics[width=0.63\columnwidth]{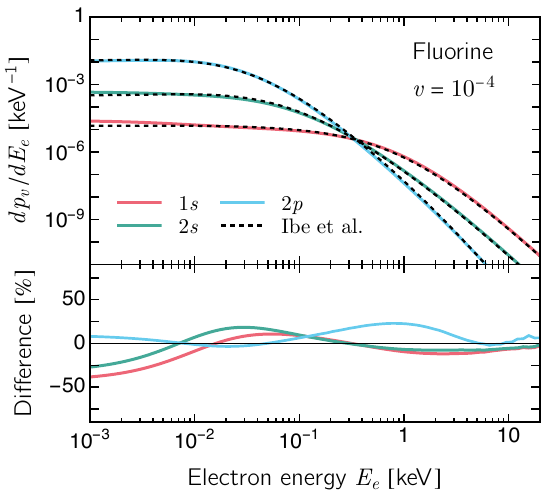}
    \includegraphics[width=0.63\columnwidth]{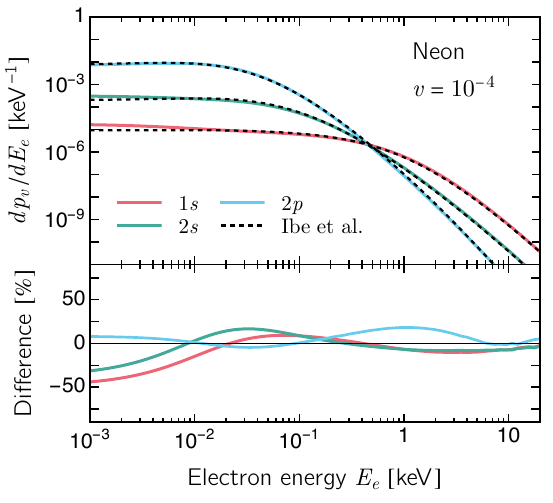}\\\vspace{1em}
    \includegraphics[width=0.63\columnwidth]{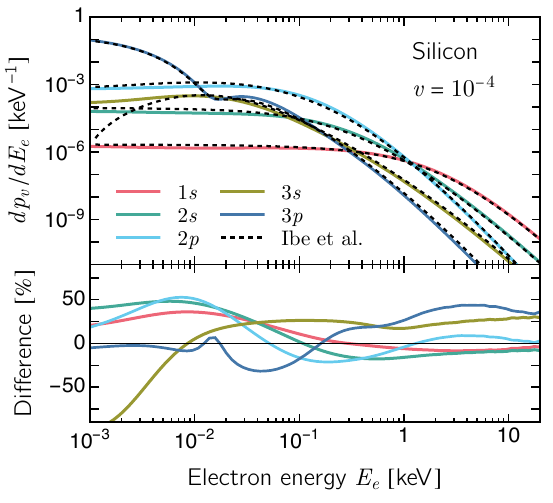}
    \includegraphics[width=0.63\columnwidth]{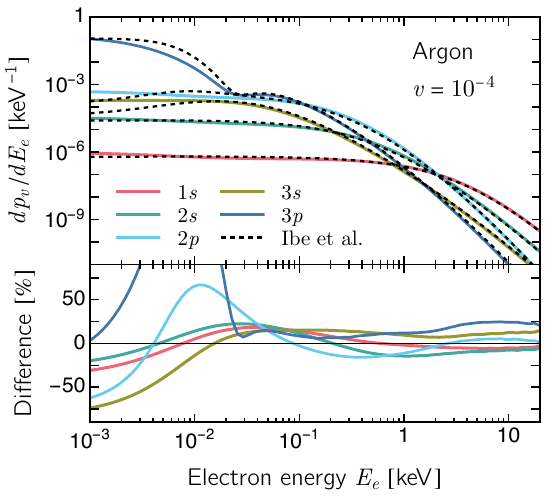}
    \includegraphics[width=0.63\columnwidth]{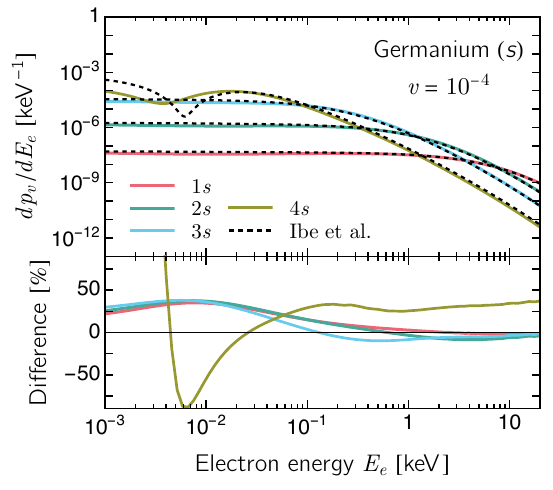}\\\vspace{1em}
    \includegraphics[width=0.63\columnwidth]{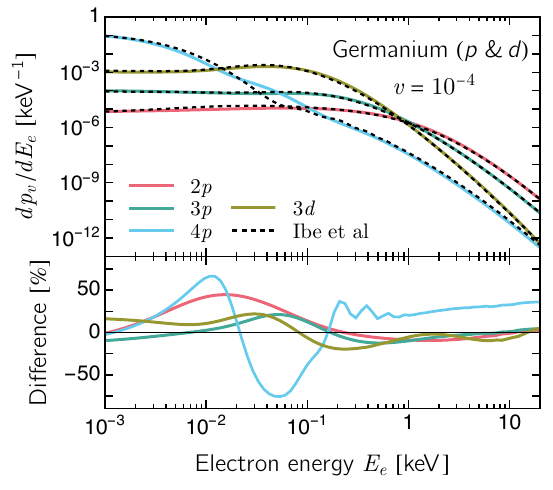}
    \includegraphics[width=0.63\columnwidth]{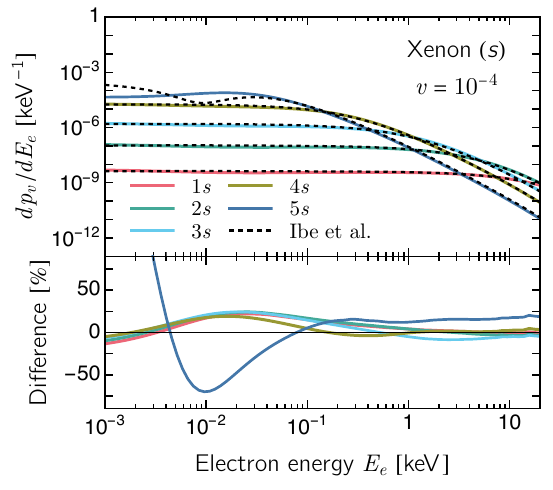}
    \includegraphics[width=0.63\columnwidth]{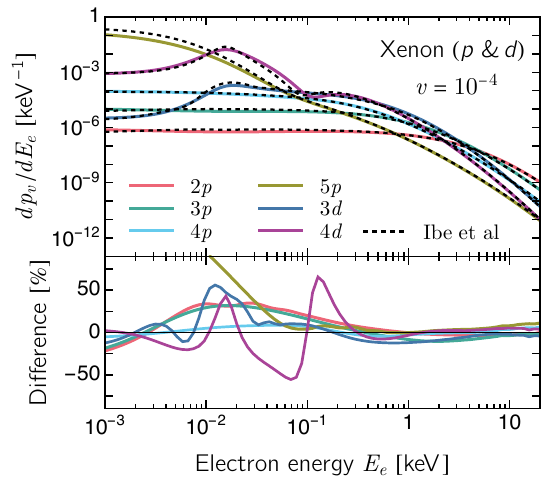}
    \caption{The upper part of each panel compares our results (coloured lines) for the differential single ionisation probability, $dp_v/dE_e$, with the equivalent results from Ibe et al.~\cite{Ibe:2017yqa} (dashed black lines). Following Ref.~\cite{Ibe:2017yqa}, we label the states with the non-relativistic quantum numbers and, for clarity, we have separated the $s$ and $p, d$ states for germanium and xenon into separate panels. The lower part of each panel shows the difference, $\left[1/(2\pi)dp^c_e/dE_e\right]/\left[dp_v/dE_e \right] -1$, expressed as a percentage.}
    \label{fig:ibe_comparison}
\end{figure*}

In this appendix we provide a comparison of our results for the differential single ionisation probability, $dp_v/dE_e$, with the results from Ref.~\cite{Ibe:2017yqa}. The dipole approximation is used in Ref.~\cite{Ibe:2017yqa}, so we make the comparison at $v=10^{-4}$ where the dipole approximation provides accurate results.

In \cref{fig:ibe_comparison}, the coloured lines in the upper part of each panel show our results for ionisation from each subshell in carbon, fluorine, neon, argon and xenon. The black dashed lines in each panel show the $1/(2\pi)dp^c/dE_e$ values from Ref.~\cite{Ibe:2017yqa}. We use a different convention to normalise the continuum spinors, so our differential probability does not require the $2\pi$ factor. Following Ref.~\cite{Ibe:2017yqa}, we label the states with the non-relativistic quantum numbers $(n,\ell)$. This means that the $p$-state probabilities are the sum of those for the $p_{1/2}$ and $p_{3/2}$ relativistic states, while the $d$-state probabilities are the sum of $d_{3/2}$ and $d_{5/2}$. The lower part of each panel shows $\left[1/(2\pi)dp^c_e/dE_e\right]/\left[dp_v/dE_e \right] -1$ expressed as a percentage.

In general, we find good agreement at the level of about 30\% or better across all of the atomic species. Larger departures occur but typically at very low values of the electron energy, $1\,\mathrm{eV}\lesssim E_e \lesssim 100\,\mathrm{eV}$, where the precise form of the potential has more of an impact. Some degree of deviation is expected since Ref.~\cite{Ibe:2017yqa} employs a relativistic self-consistent mean-field approach with a local central potential, while we use the canonical Dirac-Hartree-Fock method, which includes the full non-local exchange potential.

\section{Neutron cross-sections \label{sec:nxsecs}}

\begin{table*}[ht]
    \caption{Cross-sections (in mb) for 2.47\,MeV and 14.7\,MeV neutrons from ENDF/B-VIII.0~\cite{Brown:2018}. The cross-sections are given for the main isotopes except for neon and xenon, where the abundance of additional isotopes is substantial ($\gtrsim 5\%$), so weighted averages at natural abundance are used. $\sigma_0$ denotes the total cross-section, and the signal-inducing processes ($\sigma_s$) include elastic scattering $(n,n)$, inelastic scattering $(n,n')$, $(n,2n)$ reactions and radiative capture $(n,\gamma)$.}
    \vspace{5pt}
    \centering
    {\small
    \setlength{\tabcolsep}{6pt}
    \renewcommand{\arraystretch}{1.2}
    \begin{tabular}{l | r r r r r r  | r r r r r  r}
        \hline
        & \multicolumn{6}{c|}{$E_n$ = 2.47\,MeV (D-D neutrons)} & \multicolumn{6}{c}{$E_n$ = 14.7\,MeV (D-T neutrons)} \\
                        & $^{4}$He & $^{12}$C  & $^{19}$F & $^{\mathrm{nat}}$Ne & $^{40}$Ar & $^{\mathrm{nat}}$Xe  
                        & $^{4}$He & $^{12}$C  & $^{19}$F & $^{\mathrm{nat}}$Ne & $^{40}$Ar & $^{\mathrm{nat}}$Xe \\
        \hline
        $\sigma_0$          & 3,239 & 1,613 & 3,038 & 2,474 & 5,050$^\dagger$  & 5,760 
                            & 1,017 & 1,379 & 1,786 & 1,677 & 2,818            & 4,871 \\
        $(n,n)$             & 3,239 & 1,613 & 2,131 & 2,028 & 4,318$^\dagger$  & 3,876
                            & 1,017 &   895 &   985 &   771 & 1,556            & 2,933 \\
        $(n,n')$            &    -- &    -- &   907 &   437 &   731            & 1,869 
                            &    -- &   426 &   235 &   240 &   553            &   464 \\
        $(n,2n)$            &    -- &    -- &    -- &    -- &    --            &    -- 
                            &    -- &    -- &    52 &    44 &   645            & 1,463 \\
        $(n,\gamma)$        &    -- &  0.05 &  0.09 &  0.10 &  0.32            &    15
                            &    -- &  0.15 &  0.03 &  0.02 &  0.08            &   0.72 \\
        $\sigma_s/\sigma_0$ & 100\% & 100\% & 100\% & 100\% & 100\%            & 100\%
                            & 100\% & 96\%  &  71\% &  63\% &  98\%            & 100\% \\
    \hline
    \end{tabular}
    }
    \label{Tab:nxs}
    \begin{flushleft}
        {\footnotesize 
        $^\dagger$ These cross-sections are in the resonance region for $^{40}$Ar($n$,$n$) and vary rapidly with energy. The average cross-sections over the range from 2.45\,MeV to 2.50\,MeV are $\sigma_0=4,\!327\,\mathrm{mb}$ and $3,\!594\,\mathrm{mb}$ for $(n,n)$. We use the average values in our calculations.}
    \end{flushleft}
\end{table*}

\begin{figure*}[t!]
    \includegraphics[width=0.9\columnwidth]{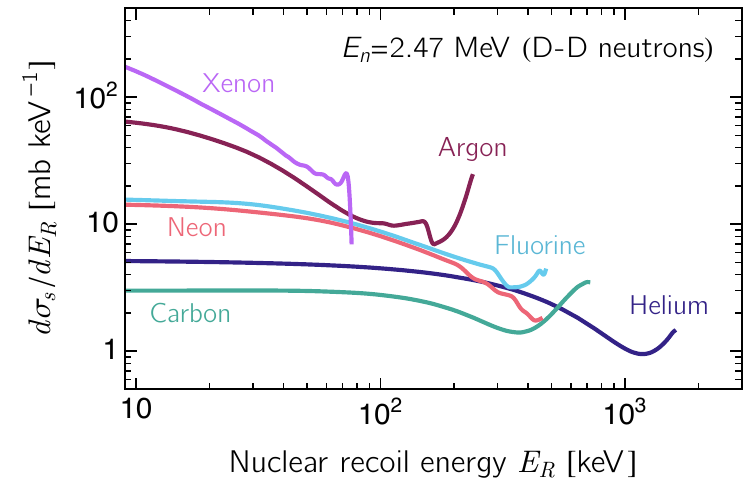}
    \hspace{1.0ex}
    \includegraphics[width=0.9\columnwidth]{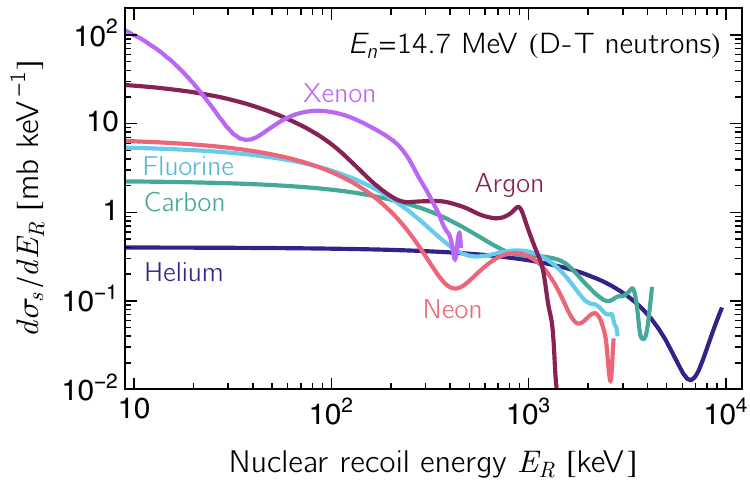}
    \caption{Differential neutron cross-sections for signal-inducing processes. The left (right) panel shows the spectra generated by a D-D (D-T) generator with neutrons at 2.47 (14.7)~MeV. Each curve is drawn to the maximum recoil energy, which decreases for heavier atoms and lower neutron energy.}
    \label{fig:neutronxsecs}
\end{figure*}

This appendix gives the neutron--nucleus cross-sections used in this work. The numerical values of the cross-sections in~\cref{Tab:nxs} are from the ENDF/B-VIII.0 library~\cite{Brown:2018}. The values on the left (right) correspond to nominal neutron energies from a D-D (D-T) neutron generator. We have listed cross-sections for elastic scattering, inelastic scattering, $(n,2n)$ reactions and radiative capture processes as all of these processes can give rise to an electron and nuclear recoil track with a common vertex: the signal for which the MIGDAL experiment is searching~\cite{MIGDALa}.

\Cref{fig:neutronxsecs} shows the combined differential cross-section for all signal-inducing processes as a function of the nuclear recoil energy. The left and right panels show the spectra expected from an incoming neutron with energy 2.47\,MeV and 14.7\,MeV, respectively, and the curves extend to the end-point recoil energies. The spectra were generated with GEANT4~v10.5.1 (G4NDL~4.5)~\cite{GEANT4:2002zbu}.


\newpage
\bibliographystyle{apsrev-4-2-titlesshown}
\bibliography{biblio}

\end{document}